\documentclass[11pt,eqs]{article}
\usepackage{latexsym}

\def\cleq{\setcounter{equation}{0}}

\makeatletter
\newcommand\xleftrightarrow[2][]{%
  \ext@arrow 9999{\longleftrightarrowfill@}{#1}{#2}}
\newcommand\longleftrightarrowfill@{%
  \arrowfill@\leftarrow\relbar\rightarrow}
\makeatother

\title{
T-dualization of type II superstring theory in double space
\thanks{Work supported in part by
the Serbian Ministry of Education, Science and Technological Development, under contract No. 171031.}}
\author{B. Nikoli\'c and B. Sazdovi\'c
\thanks{e-mail: bnikolic, sazdovic@ipb.ac.rs}\\
{\it Institute of Physics Belgrade,}\\
{\it University of Belgrade,}\\
{\it Pregrevica 118, Serbia}}

\begin{document}

\maketitle
\begin{abstract}
In this article we offer the new interpretation of T-dualization procedure of type II
superstring theory in double space framework. We use the ghost free action of type II superstring in pure spinor formulation in approximation
of constant background fields up to the quadratic terms. T-dualization along any subset of the initial coordinates, $x^a$,
is equivalent to the permutation of this subset with subset of the corresponding T-dual coordinates, $y_a$, in double space coordinate $Z^M=(x^\mu,y_\mu)$.
Demanding that the T-dual transformation law after exchange $x^a\leftrightarrow y_a$ has the same form as initial one, we obtain the T-dual
NS-NS and NS-R background fields. The T-dual R-R field strength is determined up to one
arbitrary constant under some assumptions. The compatibility between supersymmetry and T-duality produces change of bar spinors and R-R field strength. If we dualize odd number of dimensions $x^a$, such change
flips type IIA/B to type II B/A. If we T-dualize time-like direction, one imaginary unit $i$  maps type II superstring theories to type $II^\star$ ones.
\end{abstract}
%%%%%%%%%%%%%%%%%%%%%%%%%%%%%%%%%%%%%%%%%%%%%%%%%%%%%%%%%%%%%%%%%%%%%%%%%%%%%%%%%%%%%

\section{Introduction}
\setcounter{equation}{0}

T-duality is a fundamental feature of string theory \cite{S,B,RV,GPR,AABL}. As a consequence of T-duality  there is no physical
difference between string theory compactified on a circle of radius $R$ and circle of radius $1/R$. This conclusion can be generalized to tori of various dimensions.

Mathematical realization of T-duality is given by Buscher T-dualization procedure \cite{B}.
If the background fields have global isometries along some
directions then we can localize that symmetry introducing gauge fields.
The next step is to add the new term in the
action with Lagrange multipliers which forces these gauge fields to be unphysical.
Finally, we can use gauge freedom to fix initial coordinates.
Varying this gauge fixed action with respect to the Lagrange
multipliers one gets the initial action and varying with respect to the gauge fields one gets T-dual action.

Buscher T-dualization can be applied along directions
on which background fields do not depend \cite{B,RV,GPR,AABL,nasnpb,englezi}. Such procedure was used in the papers \cite{Lust,ALLP,ALLP2,L,nongeo1} in the context
of closed string noncommutativity. There is a generalized Buscher procedure which deals with background fields depending on all coordinates.
The generalized procedure was applied to the case of bosonic string moving in the weakly curved background \cite{DS1,DNS2}. It leads directly to closed string
noncommutativity \cite{DNS}.

The Buscher procedure can be considered as definition of T-dualization.
But there are also other frameworks
in which we can represent T-dualization which must be in accordance with the Buscher procedure.
Here we speak about double space formalism which was the subject of the articles about twenty years
ago \cite{Duff,AAT1,AAT2,WS1,WS2}. Double space is spanned by coordinates $Z^M=(x^\mu,y_\mu)$ $(\mu=0,1,2,\dots,D-1)$,
where $x^\mu$ and $y_\mu$ are the coordinates of the $D$-dimensional initial and T-dual space-time, respectively.
Interest for this subject emerged again with papers \cite{Hull,Hull2,berman,negeom,hohmz},
where T-duality is related with $O(d,d)$ transformations.
The approach of Ref.\cite{Duff} has been recently improved when
the T-dualization along some subset of the initial and corresponding subset of the T-dual coordinates has been
interpreted as permutation of these subsets in the double space coordinates \cite{sazdam,sazda}.

Let us motivate our interest for this subject. It is well known that T-duality is important feature  in understanding the M-theory. In fact, five consistent  superstring theories are connected by web of T and S dualities. 
In the beginning we are going to pay attention to the T-duality.  
To  obtain formulation of M-theory it is not enough  to find all corresponding T-dual theories. We must construct one theory which contain the initial theory and all corresponding T-dual ones.

We have succeeded to realize such program in the bosonic case, for  both  constant and  weakly curved background.  
In Refs.\cite{sazdam,sazda} we doubled all bosonic coordinates and obtain the theory which contains the initial and all corresponding T-dual theories. 
In such theory T-dualization along arbitrary set of coordinates $x^a$ is equivalent to 
replacement of these coordinates with corresponding T-dual ones $y_a$. 
So,  T-duality in double space becomes  symmetry transformation with respect to  permutation group.
 
Performing T-duality in supersymmetric case generates new problems. In the present paper  we are going to extend such approach to the type II theories. 
In fact, doubling all bosonic coordinates we have unified types IIA, IIB as well as type $II^\star$ \cite{timelike} (obtained by T-dualization along time-like direction)  theories. 
We expect that such a program could be a step toward better understanding M-theory.

In the present article we apply the approach of Refs.\cite{sazdam,sazda} in the cases of complete (along all bosonic coordinates) and partial
(subset of the bosonic coordinates) T-dualization of the type II superstring theory \cite{S}. We use ghost free type II superstring theory in pure spinor
formulation \cite{berko,susyNC,NPBref} in the approximation of constant background fields and up to the quadratic terms. This action is obtained from the general type II superstring action 
\cite{verteks} which is given in the form of an expansion in powers of fermionic coordinates $\theta^\alpha$ and $\bar\theta^\alpha$. In the first step of consideration
we will limit our analysis on the basic term of the action neglecting $\theta^\alpha$ and $\bar\theta^\alpha$ dependent terms. Later, in the discussion of proper fermionic variables,
using iterative procedure \cite{verteks}, we take into consideration higher power terms and restore supersymmetric invariants $\Pi_\pm^\mu$, $d_\alpha$ and $\bar d_\alpha$ as variables in the theory.

Rewriting the T-dual transformation laws
in terms of the double space coordinates $Z^M$ we introduce the generalized metric ${\cal H}_{MN}$ and the generalized current $J_{\pm M}$.  The permutation matrix $({\cal T}^a)^M{}_N$
exchanges the places of $x^a$ and $y_a$, where index $a$ marks the directions along which we make T-dualization. The basic
demand is that T-dual double space coordinates, ${}_a Z^M=({\cal T}^a)^M{}_N Z^N$, satisfy the transformation law of the same form as initial coordinates, $Z^M$.
It produces the expressions for T-dual generalized metric, ${}_a {\cal H}_{MN}=({\cal T}^a{\cal H}{\cal T}^a)_{MN}$, and T-dual current, ${}_a J_{\pm M}=({\cal T}^a J_{\pm})_M$.
This is equivalent to the requirement that transformations of the coordinates and background fields, $Z^M\to {}_a Z^M$, ${\cal H}_{MN}\to {}_a{\cal H}_{MN}$ and $J_{\pm M}\to {}_a J_{\pm M}$, are symmetry transformations of the double space action.
From transformation of the generalized metric we obtain T-dual NS-NS background fields and from
transformation of the current we obtain T-dual NS-R fields.

The supersymmetry case includes the new features in both Buscher and double space T-duality approaches. In the bosonic case the left and right world-sheet
chiralities have different T-duality transformations. It implies that in T-dual theory two fermionic coordinates, $\theta^\alpha$ and $\bar\theta^\alpha$, and corresponding
canonically conjugated momenta, $\pi_\alpha$ and $\bar\pi_\alpha$ (with different world-sheet chiralities), have different supersymmetry transformations.
As it is shown in \cite{H,BPT} it is possible to make supersymmetry transformation in T-dual theory unique if we change one world-sheet chirality sector. So, compatibility between
supersymmetry and T-duality can be achieved by action on bar variables with operator ${}_a\Omega$, ${}^\bullet \bar \pi_\alpha={}_a\Omega_\alpha{}^\beta \;{}_a\bar \pi_\beta$. As a consequence
of the relation $\Gamma^{11}\;{}_a\Omega=(-1)^d \;{}_a\Omega \Gamma^{11}$ it follows that such transformations for odd $d$ change space-time chiralities of the bar spinors. In such a way
operator ${}_a\Omega$ for odd $d$ maps type IIA/B to type IIB/A theory. Here $d$ denotes the number of T-dualized directions.

There is one difference comparing with bosonic string case \cite{sazdam,sazda} where all results from Buscher
procedure were reproduced. In the T-dual transformation laws of type II superstring theory the R-R field strength $F^{\alpha\beta}$ does not appear.
The reason is that R-R field strength couples only with the fermionic degrees of freedom which are not dualized. This is analogy with the term $\partial_+ x^i \Pi_{+ ij}\partial_- x^j$ in the bosonic case, where background field $\Pi_{+ij}$ couples only with coordinates $x^i$ which are undualized \cite{Hull,Hull2}.
To reproduce Buscher form of the T-dual R-R field strength we
should make some additional assumptions.

There is one appendix which contains block-wise expressions for tensors used in this article and useful relations.
%%%%%%%%%%%%%%%%%%%%%%%%%%%%%%%%%%%%%%%%%%%%%%%%%%%%%%%%%%%%%%%%%%%%%%%%%%%%%%%%%%%%%

\section{Buscher T-dualization of type II superstring theory}
\setcounter{equation}{0}

In this section we will consider type II superstring action in pure spinor formulation \cite{berko,susyNC,NPBref}
in the approximation of constant background fields and up to the quadratic terms. Then we will give the
overview of the results obtained by Buscher T-dualization procedure \cite{H,BPT,nasnpb,englezi}.

\subsection{Type II superstring in pure spinor formulation}

The sigma model action for type II superstring of
Ref.\cite{verteks} is of the form
\begin{equation}\label{eq:vsg}
S=\int_\Sigma d^2 \xi (X^T)^M A_{MN}\bar X^N+S_\lambda+S_{\bar\lambda}\, ,
\end{equation}
where vectors $X^M$ and $\bar X^N$ are left and right chiral supersymetric variables
\begin{equation}
X^M=\left(
\begin{array}{c}
\partial_+ \theta^\alpha\\\Pi_+^\mu\\d_\alpha\\\frac{1}{2}N_+^{\mu\nu}
\end{array}\right)\, ,\quad \bar X^M=\left(
\begin{array}{c}
\partial_-\bar\theta^\alpha\\\Pi_-^\mu\\\bar d_\alpha\\\frac{1}{2}\bar N_-^{\mu\nu}
\end{array}\right),
\end{equation}
which components are defined as
\begin{equation}
\Pi_+^\mu=\partial_+ x^\mu+\frac{1}{2}\theta^\alpha (\Gamma^\mu)_{\alpha\beta}\partial_+ \theta^\beta\, ,\quad \Pi_-^\mu=\partial_- x^\mu+\frac{1}{2}\bar\theta^\alpha (\Gamma^\mu)_{\alpha\beta}\partial_- \bar\theta^\beta\, ,
\end{equation}
\begin{eqnarray}
d_\alpha&=&\pi_\alpha-\frac{1}{2}(\Gamma_\mu \theta)_\alpha\left[ \partial_+ x^\mu +\frac{1}{4} (\theta \Gamma_\mu \partial_+\theta)\right]\, ,\nonumber\\
\bar d_\alpha&=&\bar\pi_\alpha-\frac{1}{2}(\Gamma_\mu \bar\theta)_\alpha \left[\partial_- x^\mu +\frac{1}{4} (\bar\theta \Gamma_\mu \partial_-\bar\theta)\right]\, ,
\end{eqnarray}
\begin{equation}\label{eq:Npm}
N_+^{\mu\nu}=\frac{1}{2}w_\alpha(\Gamma^{[\mu\nu]})^\alpha{}_\beta \lambda^\beta\, ,\quad \bar N_-^{\mu\nu}=\frac{1}{2}\bar w_\alpha (\Gamma^{[\mu\nu]})^\alpha{}_\beta \bar\lambda^\beta\, .
\end{equation}
Inserting the supermatrix $A_{MN}$
\begin{equation}\label{eq:Amn}
A_{MN}=\left(\begin{array}{cccc}
A_{\alpha\beta} & A_{\alpha\nu} & E_\alpha{}^\beta & \Omega_{\alpha,\mu\nu}\\
A_{\mu\beta} & A_{\mu\nu} & \bar E_\mu^\beta & \Omega_{\mu,\nu\rho}\\
E^\alpha{}_\beta & E^\alpha_\nu & {\rm P}^{\alpha\beta} & C^\alpha{}_{\mu\nu}\\
\Omega_{\mu\nu,\beta} & \Omega_{\mu\nu,\rho} & \bar C_{\mu\nu}{}^\beta & S_{\mu\nu,\rho\sigma}
\end{array}\right)\, ,
\end{equation}
in (\ref{eq:vsg}), action gets the expanded form \cite{verteks}
\begin{eqnarray}\label{eq:VSG}
S&=&\int d^2\xi \left[\partial_+ \theta^\alpha A_{\alpha\beta}\partial_-\bar\theta^\beta+\partial_+ \theta^\alpha A_{\alpha\mu}\Pi_-^\mu+\Pi_+^\mu A_{\mu\alpha}\partial_-\bar\theta^\alpha+\Pi_+^\mu A_{\mu\nu}\Pi_-^\nu\right.\nonumber\\
&+& d_\alpha E^\alpha{}_\beta \partial_-\bar\theta^\beta+d_\alpha E^\alpha{}_\mu \Pi_-^\mu+\partial_+ \theta^\alpha E_\alpha{}^\beta \bar d_\beta+\Pi_+^\mu E_\mu{}^\beta \bar d_\beta+d_\alpha {\rm P}^{\alpha\beta}\bar d_\beta\nonumber\\
&+& \frac{1}{2}N_+^{\mu\nu} \Omega_{\mu\nu,\beta} \partial_- \bar\theta^\beta+\frac{1}{2}N_+^{\mu\nu} \Omega_{\mu\nu,\rho} \Pi_-^\rho+\frac{1}{2}\partial_+\theta^\alpha \Omega_{\alpha,\mu\nu}\bar N_-^{\mu\nu}+\frac{1}{2}\Pi_+^\mu \Omega_{\mu,\nu\rho}\bar N_-^{\nu\rho}\nonumber\\
&+&\left. \frac{1}{2}N_+^{\mu\nu}\bar C_{\mu\nu}{}^\beta \bar d_\beta+\frac{1}{2}d_\alpha C^\alpha{}_{\mu\nu} \bar N_-^{\mu\nu}+\frac{1}{4}N_+^{\mu\nu} S_{\mu\nu,\rho\sigma}\bar N_-^{\rho\sigma}\right]+S_\lambda+S_{\bar\lambda}\, .
\end{eqnarray}

The world sheet $\Sigma$ is parameterized by
$\xi^m=(\xi^0=\tau\, ,\xi^1=\sigma)$ and
$\partial_\pm=\partial_\tau\pm\partial_\sigma$. Superspace is spanned by bosonic coordinates $x^\mu$ ($\mu=0,1,2,\dots,9$) and fermionic ones $\theta^\alpha$ and $\bar\theta^{\alpha}$
$(\alpha=1,2,\dots,16)$. The variables $\pi_\alpha$ and $\bar
\pi_{\alpha}$ are canonically conjugated momenta to
$\theta^\alpha$ and $\bar\theta^\alpha$, respectively. The actions for pure spinors, $S_\lambda$ and $S_{\bar\lambda}$, are free field actions
\begin{equation}
S_\lambda=\int d^2\xi w_\alpha \partial_-\lambda^\alpha\, ,\quad S_{\bar\lambda}=\int d^2\xi \bar w_\alpha \partial_+ \bar\lambda^\alpha\, ,
\end{equation}
where $\lambda^\alpha$ and $\bar\lambda^\alpha$ are pure spinors and $w_\alpha$ and $\bar w_\alpha$ are their canonically conjugated momenta, respectively. The pure spinors satisfy so called pure spinor constraints
\begin{equation}\label{eq:psc0}
\lambda^\alpha (\Gamma^\mu)_{\alpha\beta}\lambda^\beta=\bar\lambda^\alpha (\Gamma^\mu)_{\alpha\beta}\bar\lambda^\beta=0\, .
\end{equation}

Matrix $A_{MN}$ containing type II superfields generally depends on $x^\mu$,
$\theta^\alpha$ and $\bar\theta^\alpha$.
The superfields $A_{\mu\nu}$, $\bar E_\mu{}^\alpha$, $E^\alpha{}_\mu$ and ${\rm P}^{\alpha\beta}$ are physical superfields, because their first components are supergravity fields.
The fields in the first column and first row are auxiliary superfirlds because they can be expressed in
terms of the physical ones \cite{verteks}. The rest ones, $\Omega_{\mu,\nu\rho}(\Omega_{\mu\nu,\rho})$, $C^\alpha{}_{\mu\nu}(\bar C_{\mu\nu}{}^\alpha)$ and $S_{\mu\nu,\rho\sigma}$, are curvatures (field strengths) for physical superfields.

The action from which we start (\ref{eq:VSG}) could be considered as an expansion in powers of $\theta^\alpha$ and $\bar\theta^\alpha$. In an iterative procedure presented in \cite{verteks} 
it has been shown that each component in the expansion
can be obtained from the previous one. So, for practical reasons (computational simplicity), in the first step we limit our considerations on the basic component i.e. we neglect all terms 
in the action containing $\theta^\alpha$ and $\bar\theta^\alpha$. As a consequence $\theta^\alpha$ and $\bar\theta^\alpha$ terms disappear from $\Pi_\pm^\mu$, $d_\alpha$ and $\bar d_\alpha$ and in the solutions for 
physical superfields just $x$-dependent supergravity fields survive. 
Therefore we lose explicit supersymmetry in such approximation. Later, when we discuss proper fermionic variables, we would go further in the expansion and take higher power terms, which means that supersymmetric invariants, $\Pi_\pm^\mu$, $d_\alpha$ and $\bar d_\alpha$, would take the 
roles of $\partial_\pm x^\mu$, $\pi_\alpha$ and $\bar \pi_\alpha$, respectively.

We are going to perform T-dualization along some subset of bosonic coordinates $x^a$. So, we will assume that these directions are Killing vectors. Since $\partial_\pm x^a$ appears in $\Pi_\pm^\mu$, $d_\alpha$ 
and $\bar d_\alpha$,
it essentially means that corresponding superfields ($A_{ab}$, $\bar E_a{}^\alpha$, $E^\alpha{}_a$, ${\rm P}^{\alpha\beta}$) should not depend on $x^a$. 
This assumption regarding Killing spinors could be extended on 
all space-time directions $x^\mu$ which effectivelly means, in the first step, that physical superfields are constant.  All auxiliary superfields can be expressed in terms of space-time derivatives of physical supergravity
fields \cite{verteks}. Then, in the first step, auxiliary superfields are zero, because all physical superfields are constant.
On the other hand, constant physical superfields means that their field strengths, 
$\Omega_{\mu,\nu\rho}(\Omega_{\mu\nu,\rho})$, $C^\alpha{}_{\mu\nu}(\bar C_{\mu\nu}{}^\alpha)$ and $S_{\mu\nu,\rho\sigma}$, are zero. In this way, in the first step, we eliminated from the action terms containing variables $N_+^{\mu\nu}$ and $\bar N_-^{\mu\nu}$ (\ref{eq:Npm}). 

This choice of background fields should be discussed from the viewpoint of space-time
field equations of type II superstring action \cite{dufftasi}.
Let us pay attention on the space-time field equations for type II superstring given in Appendix B of \cite{dufftasi}. The equation (B.7) from this set of equations represents the backreaction
of ${\rm P}^{\alpha\beta}$ on the metric $G_{\mu\nu}$. If we take constant dilaton $\Phi$ and constant antisymmetric NS-NS field
$B_{\mu\nu}$ we obtain that
\begin{equation}
R_{\mu\nu}-\frac{1}{2}G_{\mu\nu} R\sim ({\rm P}^{\alpha\beta})^2_{\mu\nu}\, .
\end{equation}
If we choose the background field ${\rm P}^{\alpha\beta}$ to be constant, in general, we will have constant Ricci tensor which means that metric tensor is quadratic function of space-time coordinates i.e. there is
back-reaction of R-R field strength on metric tensor. If one wants to cancel non-quadratic terms originating from back-reaction, additional conditions must be imposed on R-R
field strength - $AdS_5 \times S_5$ coset geometry or self-duality condition (see the first reference in \cite{berko}).

%But if we assume that
%${\rm P}^{\alpha\beta}$ is infinitesimal, then we can take $G_{\mu\nu}$ to be constant but in linear approximation in ${\rm P}^{\alpha\beta}$. This approximation could be understood as a limit of validity of our approximation of 
%constant background fields. It is clear that our choice of background fields, in general, does not obey space-time field equations, but they are obeyed approximately. Consequently, all considerations and conclusions
%made using quadratic action are of the limited validity originating from space-time field equations. 

%It remains to discuss the background fields $\bar E_\mu{}^\alpha$ and $E^\alpha{}_\mu$.  
%Taking NS-NS and R-R fields to be constant, using the equation (5.7) from \cite{verteks}, we get 
%\begin{equation}
%\bar E_\mu{}^\alpha=\bar \Psi^\alpha_\mu+(\Gamma_\mu \theta)_\gamma f^{\gamma\alpha}\, ,\quad E^\alpha{}_\mu=\Psi^\alpha_\mu+(\Gamma_\mu \bar\theta)_\gamma f^{\alpha\gamma}\, .
%\end{equation}
%The quadratic approximation demands that we drop $\theta$ and $\bar\theta$ dependent terms.

Taking into account above analysis and arguments, our approximation can be realized in the following way
\begin{equation}
\Pi_\pm^\mu\to \partial_{\pm} x^\mu\, ,\quad d_\alpha \to \pi_\alpha\, ,\quad \bar d_\alpha\to \bar\pi_\alpha\, ,
\end{equation}
and physical superfields take the form
\begin{equation}
A_{\mu\nu}=\kappa(\frac{1}{2}G_{\mu\nu}+B_{\mu\nu})\, ,\quad E^\alpha_\nu=-\Psi^\alpha_\nu\, ,\quad \bar E_\mu^\alpha=\bar\Psi_\mu^\alpha\, ,\quad {\rm P}^{\alpha\beta}=\frac{2}{\kappa}P^{\alpha\beta}=\frac{2}{\kappa}e^{\frac{\Phi}{2}}F^{\alpha\beta}\, ,
\end{equation}
where $G_{\mu\nu}$ is metric tensor and $B_{\mu\nu}$ is antisymmetric NS-NS background field.
%The final form of the vertex operator is
%\begin{eqnarray}
%V_{SG}=\int_\Sigma d^2\xi \left[
%\kappa(\frac{1}{2}g_{\mu\nu}+B_{\mu\nu})\partial_+x^\mu\partial_-x^\nu-d_\alpha
%\Psi^\alpha_\mu \partial_-x^\mu+\partial_+ x^\mu
%\bar\Psi^\alpha_\mu\bar\pi_\alpha+\frac{2}{\kappa}\pi_\alpha
%P^{\alpha\beta}\pi_\beta\right]\, .
%\end{eqnarray}
Consequently, the full action $S$ is
\begin{eqnarray}\label{eq:SB}
&{}&S=\kappa \int_\Sigma d^2\xi \left[\partial_{+}x^\mu
\Pi_{+\mu\nu}\partial_- x^\nu+\frac{1}{4\pi\kappa}\Phi R^{(2)}\right] \\&+&\int_\Sigma d^2 \xi \left[
-\pi_\alpha
\partial_-(\theta^\alpha+\Psi^\alpha_\mu
x^\mu)+\partial_+(\bar\theta^{\alpha}+\bar \Psi^{\alpha}_\mu
x^\mu)\bar \pi_{\alpha}+\frac{2}{\kappa}\pi_\alpha P^{\alpha
\beta}\bar \pi_{\beta}\right ]\, ,\nonumber
\end{eqnarray}
where
\begin{equation}\label{eq:pipm1}
\Pi_{\pm \mu\nu}=B_{\mu\nu}\pm \frac{1}{2}G_{\mu\nu}\, .
\end{equation}
Actions $S_\lambda$ and $S_{\bar\lambda}$ are decoupled from the rest and can be neglected in the further analysis. The action, in its final form, is  ghost independent.

NS-NS sector of the theory described by (\ref{eq:SB}) contains gravitation $G_{\mu\nu}$, antisymmetric
Kalb-Ramond field $B_{\mu\nu}$ and dilaton field $\Phi$. In NS-R sector there are two gravitino fields
$\Psi^\alpha_\mu$ and $\bar\Psi^\alpha_\mu$ which are Majorana-Weyl spinors of the opposite chirality in type IIA and same chirality in type IIB theory. The field $F^{\alpha\beta}$ is R-R field strength and can be expressed in terms of the antisymmetric tensors $F_{(k)}$ \cite{nasnpb,duf,grk,BNBSPLB}
\begin{equation}\label{eq:RRpolje}
F^{\alpha\beta}=\sum_{k=0}^D
\frac{1}{k!}F_{(k)}\Gamma_{(k)}^{\alpha\beta}\, , \quad \left[
\Gamma_{(k)}^{\alpha\beta}=(\Gamma^{[\mu_1\dots
\mu_k]})^{\alpha\beta}\right]
\end{equation}
where
\begin{equation}\label{eq:bazis}
\Gamma^{[\mu_1 \mu_2\dots \mu_k]}\equiv \Gamma^{[\mu_1}\Gamma^{\mu_2}\dots \Gamma^{\mu_k]}\, ,
\end{equation}
is completely antisymmetrized product of gamma matrices. The bispinor $F^{\alpha\beta}$ satisfies chirality
condition, $\Gamma^{11} F=\pm F\Gamma^{11}$, where $\Gamma^{11}$ is a product of gamma matrices in $D=10$ dimensional space-time and sign $+$ corresponds to type IIA while sign $-$ to type IIB superstring theory. Consequently, type IIA theory
contains only even rank tensors $F_{(k)}$, while type IIB odd rank tensors. Because of duality
relation, the independent tensors are $F_{(0)}$, $F_{(2)}$ and $F_{(4)}$ for type IIA while $F_{(1)}$, $F_{(3)}$ and
self-dual part of $F_{(5)}$ for type IIB superstring theory. Using mass-shell condition (massless
Dirac equation for $F^{\alpha\beta}$) these tensors can be solved
in terms of potentials $F_{(k)}=dA_{(k-1)}$. The factor $e^{\frac{\Phi}{2}}$ is in accordance with the
conventions adopted from \cite{ThompPhD}.

%%%%%%%%%%%%%%%%%%%%%%%%%%%%%%%%%%%%%%%%%%%%%%%%%%%%%%%%%%%%%%%%%%%%%%%%%%%%%%%%%%%%%%

\subsection{T-dualization along arbitrary number of coordinates}

Let us start with the action (\ref{eq:SB})
and apply standard T-dualization procedure \cite{B,DS1,DNS2}. It means that we localize the shift symmetry for some coordinates $x^a$. We substitute the ordinary derivatives with covariant ones, introducing gauge fields $v^a_\alpha$. Then we add the term $\frac{1}{2}y_a F^a_{+-}$ to the Lagrangian in order to force the field strength $F^a_{+-}$ to vanish and preserve equivalence between original and T-dual theories.  Finally, we fix  the gauge $x^a=0$ and obtain
\begin{eqnarray}\label{eq:SBfix}
&{}&S_{fix}(v^a_\pm, x^i,\theta^\alpha,\bar \theta^\alpha,\pi_\alpha,\bar \pi_\alpha) = \nonumber \\
&{}& \int_\Sigma d^2\xi \left[\kappa  v^a_+ \Pi_{+ab} v^b_-  + \kappa v^a_+ \Pi_{+a j}\partial_- x^j   + \kappa \partial_+ x^i \Pi_{+ib} v^b_- + \kappa \partial_+
x^i \Pi_{+ ij}\partial_- x^j+\frac{1}{4\pi}\Phi R^{(2)} \right.   \nonumber \\
&-& \pi_\alpha   \Psi^\alpha_b v^b_- + v^a_+   \bar \Psi^{\alpha}_a \bar \pi_{\alpha}
- \pi_\alpha \partial_-(\theta^\alpha+\Psi^\alpha_i x^i)+\partial_+(\bar\theta^{\alpha}+\bar \Psi^{\alpha}_i
x^i)\bar \pi_{\alpha}+\frac{1}{2\kappa} \, e^{\frac{\Phi}{2}} \, \pi_\alpha F^{\alpha \beta}\bar \pi_{\beta}                 \, \nonumber \\
&+& \left.   \frac{\kappa}{2} (v^a_+ \partial_- y_a -  v^a_- \partial_+ y_a)               \right]     \, .
\end{eqnarray}

Varying the gauge fixed action with respect to the Lagrange multipliers $y_a$ we get the solution
for gauge fields in the form
\begin{equation}\label{eq:vpx}
v^a_\pm=\partial_\pm x^a\, ,
\end{equation}
while varying with respect to the gauge fields $v^a_\pm$ we have
\begin{equation}\label{eq:lawsb}
v_\pm^a  =  -2\kappa \hat\theta_\pm^{ab}\Pi_{\mp bi} \partial_\pm x^i-\kappa \hat\theta^{ab}_{\pm}   \partial_\pm y_b
\pm 2 \hat\theta^{ab}_{\pm} \Psi^\alpha_{\pm b} \pi_{\pm \alpha} \, .
\end{equation}
Substituting $v_\pm^a$ in (\ref{eq:SBfix}) we find
\begin{eqnarray}\label{eq:SBfixd}
&{}& S_{fix} (y_a, x^i,\theta^\alpha,\bar \theta^\alpha,\pi_\alpha,\bar \pi_\alpha)  = \nonumber \\
&{}& \int_\Sigma d^2\xi \left[\frac{\kappa^2}{2}   \partial_+ y_a  \hat \theta_-^{ab}  \partial_- y_b   + \kappa^2 \partial_+ y_a \hat \theta_-^{ab}  \Pi_{+b j} \partial_- x^j   - \kappa^2 \partial_+ x^i \Pi_{+ia}
\hat \theta_-^{ab}  \partial_- y_b   +\frac{1}{4\pi}\Phi R^{(2)}\right.   \nonumber \\
&+&  \kappa \partial_+ x^i ( \Pi_{+ ij} -2\kappa  \Pi_{+ia}  \hat \theta_-^{ab} \Pi_{+b j}        )  \partial_- x^j   \nonumber \\
&-&  \pi_\alpha \partial_-(\theta^\alpha + \Psi^\alpha_i x^i  -2 \Psi^\alpha_a \hat \theta_-^{ab} \Pi_{+b j} x^j -  \Psi^\alpha_a  \hat \theta_-^{ab}  y_b   )        \nonumber \\
&+&  \partial_+ (\bar\theta^{\alpha}+\bar \Psi^{\alpha}_i x^i     + 2  \bar \Psi^\alpha_a \hat \theta_+^{ab} \Pi_{-b j} x^j + \bar \Psi^\alpha_a  \hat \theta_+^{ab}  y_b  ) \bar \pi_{\alpha}       \nonumber \\
&+&   \left. 2 \pi_\alpha \Psi^\alpha_a \hat \theta_-^{ab} \bar \Psi^\beta_b       \bar \pi_{\beta}    + \frac{1}{2\kappa} \, e^{\frac{\Phi}{2}} \, \pi_\alpha F^{\alpha \beta}\bar \pi_{\beta}\right]     \, \, .
\end{eqnarray}
Before we read the T-dual background fields,  we must express this action  in terms of the appropriate  spinor coordinates, which we will discuss in the next subsections.

Combining two solutions for gauge fields (\ref{eq:vpx}) and (\ref{eq:lawsb})  we obtain transformation law  between initial $x^a$ and T-dual coordinates $y_a$
\begin{equation}\label{eq:lawsb1}
\partial_\pm x^a\cong -2\kappa \hat\theta_\pm^{ab}\Pi_{\mp bi} \partial_\pm x^i  -\kappa \hat\theta^{ab}_{\pm}  ( \partial_\pm y_b  -  J_{\pm b}  )   \, .
\end{equation}
Its inverse is solution of the last equation  in terms of $y_a$
\begin{equation}\label{eq:lawsb2}
\partial_\pm y_a\cong -2\Pi_{\mp ab} \partial_\pm x^b-2\Pi_{\mp a i} \partial_\pm x^i+ J_{\pm a}      \, ,
\end{equation}
where we use $\cong$ to emphasize that these are T-duality relations.
Here we introduced the current $J_{\pm \mu}$ in the form
\begin{equation}\label{eq:strujaJmu}
J_{\pm \mu}=\pm\frac{2}{\kappa}\Psi^\alpha_{\pm \mu}\pi_{\pm \alpha}\, ,
\end{equation}
where
\begin{equation}
\Psi^\alpha_{+\mu}\equiv\Psi^\alpha_\mu\, ,\quad \Psi^{\alpha}_{-\mu}\equiv\bar\Psi^\alpha_\mu\, ,\quad \pi_{+\alpha}\equiv \pi_\alpha\, ,\quad \pi_{-\alpha}\equiv\bar \pi_\alpha\, ,
\end{equation}
and the expression  $\hat\theta^{ab}_\pm$ is defined in (\ref{eq:bfc}).

%%%%%%%%%%%%%%%%%%%%%%%%%%%%%%%%%%%%%%%%%%%%%%%%%%%%%%%%%%%%%%%%%%%%%%%

\subsection{Relation between left and right chirality in T-dual theory}

One can see from (\ref{eq:lawsb1}) and (\ref{eq:lawsb2}) that left and right chiralities transform differently in T-dual theory. As a consequence,  in T-dual theory we will have two types of vielbeins,  two types of
$\Gamma$-matrices, two  types of spin connections and two  types of supersymmetry transformations. We want to have the single geometry in T-dual theory. So, we will show that all these different representations of the same variables can be connected
by Lorentz transformations \cite{H,BPT}.

\subsubsection{Two sets of  vielbeins in T-dual theory}

The T-dual transformations of the coordinates (\ref{eq:lawsb2})  we can put in the form
\begin{equation}\label{eq:ctd}
 \left (
\begin{array}{c}
\partial_\pm y_a          \\
 \partial_\pm x^i
\end{array}\right )
=
 \left (
\begin{array}{cc}
-2 \Pi_{\mp ab}    &    -2 \Pi_{\mp aj}         \\
   0       &  \delta^i_j
\end{array}\right ) \left ( \begin{array}{c}
\partial_\pm x^b          \\
 \partial_\pm x^j
\end{array}\right ) + \left ( \begin{array}{c}
J_{\pm a}        \\
 0
\end{array}\right )  \, ,
\end{equation}
which can be rewritten as
\begin{equation}\label{eq:tdm}
\partial_+ ({}_a X)_{\hat \mu} = ({\bar Q}^{-1 T})_{{\hat \mu} \nu}  \partial_+ x^\nu  + J_{+ \hat \mu}\, ,\quad
\partial_- \, ({}_a X)_{\hat \mu} = ( Q^{-1 T})_{{\hat \mu} \nu}  \partial_- x^\nu  + J_{- \hat \mu}\, ,
\end{equation}
where we introduced the T-dual variables ${}_a X_{\hat \mu} = \{y_a , \, x^i \}$.
Here and further on the left subscript $a$ denotes the T-dualization along $x^a$ directions.
For coordinates which contain both $x^i$ and $y_a$ we will use "hat" indices $\hat \mu, \hat \nu$.   The matrices
\begin{equation}\label{eq:qbarq}
 Q^{\hat \mu \nu}  =
 \left (
\begin{array}{cc}
\kappa {\hat \theta}^{ab}_+   &    0        \\
  -2 \kappa \Pi_{- ic}  {\hat \theta}^{cb}_+        &  \delta^i_j
\end{array}\right )  \, , \qquad
\bar Q^{\hat \mu \nu}  =
 \left (
\begin{array}{cc}
\kappa {\hat \theta}^{ab}_-   &    0        \\
  -2 \kappa \Pi_{+ ic}  {\hat \theta}^{cb}_-        &  \delta^i_j
\end{array}\right ) \, ,
\end{equation}
and theirs inverse
\begin{equation}\label{eq:qbarqi}
 Q^{-1}_{\mu \hat \nu}  =
 \left (
\begin{array}{cc}
2 \Pi_{- ab}  &    0        \\
 2 \Pi_{- ib}       &  \delta^j_i
\end{array}\right )  \, , \qquad
\bar Q^{-1}_{\mu \hat \nu}  =
 \left (
\begin{array}{cc}
2 \Pi_{+ ab}  &    0        \\
 2 \Pi_{+ ib}       &  \delta^j_i
\end{array}\right ) \,  ,
\end{equation}
perform T-dualization for vector indices.

Note that different chiralities transform with different matrices $Q^{\hat \mu \nu}$ and $\bar Q^{\hat \mu \nu}$. So, there are two types of T-dual vielbeins
\begin{equation}\label{eq:tde}
{}_a e^{\underline{a} \hat \mu} = e^{\underline{a}}{}_\nu (Q^T)^{\nu \hat \mu} \, , \qquad
{}_a \bar e^{\underline{a} \hat \mu} = e^{\underline{a}}{}_\nu (\bar Q^T)^{\nu \hat \mu} \, ,
\end{equation}
with the same T-dual metric
\begin{equation}\label{eq:tdg}
 {}_a G^{\hat \mu \hat \nu} \equiv \, ({}_a e^T \eta \,{}_a e)^{\hat \mu \hat \nu} = (Q G Q^T)^{\hat \mu \hat \nu} =
 {}_a \bar G^{\hat \mu \hat \nu} \equiv \, ({}_a \bar e^T \eta \,{}_a \bar e)^{\hat \mu \hat \nu} = (\bar Q G \bar Q^T)^{\bar \mu \hat \nu}\, .
\end{equation}
The Lorentz indices are underlined (denoted by $\underline{a}, \underline{b}$).

The two T-dual vielbeins are equivalent because they are related by particular local Lorentz transformation
\begin{equation}\label{eq:lLt}
{}_a \bar e^{\underline{a} \hat \mu} = \Lambda^{\underline{a}}{}_{\underline{b}}  \, {}_a e^{\underline{b} \hat \mu}   \, , \qquad
\Lambda^{\underline{a}}{}_{\underline{b}}  = e^{\underline{a}}{}_\mu (Q^{-1} \bar Q)^{T \mu}{}_\nu  (e^{-1})^\nu{}_{\underline{b}}  \, .
\end{equation}
From (\ref{eq:qbarq}) and (\ref{eq:qbarqi}) we have
\begin{equation}\label{qm1bq}
 (Q^{-1} \bar Q)^{T \mu}{}_{\nu}  =
 \left (
\begin{array}{cc}
\delta^a{}_b + 2 \kappa \hat \theta^{ac}_+ G_{cb}   &      2 \kappa \hat \theta^{ac}_+ G_{cj}      \\
0     &  \delta^i_j
\end{array}\right )  \,  ,
\end{equation}
which produces
\begin{equation}\label{eq:lab}
\Lambda^{\underline{a}}{}_{\underline{b}} = \delta^{\underline{a}}{}_{\underline{b}} -2 \omega^{\underline{a}}{}_{\underline{b}}    \, , \qquad
\omega^{\underline{a}}{}_{\underline{b}}  = - \kappa e^{\underline{a}}{}_a  {\hat \theta}^{ab}_+  (e^T)_b{}^{\underline{c}} \, \eta_{\underline{c} \underline{b}}  \, .
\end{equation}
It satisfies  definition of Lorentz transformations
\begin{equation}\label{eq:dlt}
\Lambda^T \eta  \Lambda = \eta  \quad  \Longrightarrow  \quad  \det \Lambda^{\underline{a}}{}_{\underline{b}} = \pm 1 \, .
\end{equation}
After careful calculations we have $\det \Lambda^{\underline{a}}{}_{\underline{b}} = (-1)^d$,
where $d$ is the number of dimensions along which we perform T-duality.

\subsubsection{Two sets of $\Gamma$-matrices in T-dual theory}

Because in T-dual theory there are  two vielbeins, it must also be two sets of  $\Gamma$-matrices in curved space
\begin{equation}\label{eq:gamam1}
{}_a \Gamma_{\hat \mu} =  ({}_a e^{-1})_{ \hat \mu \underline{a}} \,   \Gamma^{\underline{a}} = ({}_a e^{-1} \Gamma)_{\hat \mu}  \, , \qquad
{}_a \bar \Gamma_{\hat \mu} =  ({}_a \bar e^{-1})_{ \hat \mu \underline{a}} \,   \Gamma^{\underline{a}}= ({}_a \bar e^{-1}  \Gamma)_{\hat \mu}   \, .
\end{equation}
They are related by the expression
\begin{equation}\label{eq:rgamam}
{}_a \bar \Gamma_{\hat \mu}  = {}_a \Omega^{-1} \, {}_a \Gamma_{\hat \mu} \; {}_a \Omega    \, ,
\end{equation}
where ${}_a \Omega$ is spinorial representation of the Lorentz transformation
\begin{equation}\label{eq:omega}
{}_a \Omega^{-1} \, \Gamma^{\underline{a}} \, \, {}_a \Omega = (\Lambda^{-1})^{\underline{a}}{}_{\underline{b}} \, \Gamma^{\underline{b}}    \, .
\end{equation}

\subsubsection{Two sets of spin connections in T-dual theory}

The spin connection can be expressed in terms of vielbein as
\begin{equation}\label{eq:spinc}
\omega_\mu {}^{\underline{a} \underline{b}} = \frac{1}{2} (e^{\nu \underline{a}} c^{\underline{b}}_{\mu \nu} - e^{\nu \underline{b}} c^{\underline{a}}_{\mu \nu}   )  -
\frac{1}{2}   e^{\rho \underline{a}} e^{\sigma  \underline{b}} c_{\underline{c} \rho \sigma} e^{\underline{c}}{}_\mu     \, ,
\end{equation}
where
\begin{equation}\label{eq:camn}
c^{\underline{a}}_{\mu \nu} = \partial_\mu  e^{\underline{a}}{}_\nu -  \partial_\nu  e^{\underline{a}}{}_\mu          \, .
\end{equation}
So, in T-dual theory there are two spin connections, defined in terms of two vielbeins. As a consequence of  (\ref{eq:lLt}) they are related as
\begin{equation}\label{eq:omegar}
{}_a {\bar \omega}^{\hat \mu} {}^{\underline{a}}{}_{\underline{b}} =  \Lambda{}^{\underline{a}}{}_{\underline{c}} \,\, {}_a  \omega{}^{\hat \mu \underline{c}}{}_{\underline{d}} \,  (\Lambda^{-1}){}^{\underline{d}}{}_{\underline{b}}
+ \Lambda{}^{\underline{a}}{}_{\underline{c}} \,  \partial^{\hat \mu} \, (\Lambda^{-1}){}^{\underline{c}}{}_{\underline{b}}             \, .
\end{equation}

It is useful to introduce the spin connection in the form
\begin{equation}\label{eq:scgg}
\omega_\mu = \omega_\mu {}_{\underline{a} \underline{b}}   \Gamma^{\underline{a} \underline{b}}          \, ,
\end{equation}
where
\begin{equation}\label{eq:gamam2}
 \Gamma^{\underline{a} \underline{b}} = \Gamma^{\underline{a}}  \Gamma^{\underline{b}}  - \Gamma^{\underline{b}}   \Gamma^{\underline{a}}         \, .
\end{equation}
Then from  (\ref{eq:omega}) for ${}_a \Omega =const$  we obtain
\begin{equation}\label{eq:oco}
{}_a \bar\omega^{\hat \mu} =   {}_a \Omega^{-1} \, {}_a \omega^{\hat \mu}  \, \, {}_a \Omega         \, .
\end{equation}

\subsubsection{Single form of supersymmetry invariants in T-dual theory and new spinor coordinates}

So far we use the action from Ref.\cite{verteks} which is an expansion in powers of $\theta^\alpha$ and $\bar\theta^\alpha$. We performed the procedure of bosonic T-dualization using first term in the expansion i.e. 
$\theta^\alpha$ and $\bar\theta^\alpha$ independent part of the action. Consequently, supersymmetric invariants, $\Pi_\pm^\mu$, $d_\alpha$ and $\bar d_\alpha$, in that approximation became $\partial_\pm x^\mu$, 
$\pi_\alpha$ and $\bar\pi_\alpha$. But if we would take higher power terms into consideration, then these invariants would appear again in the theory. Consequently, we can use these invariants 
to find proper spinor variables.

From compatibility between supersymmetry and T-duality we will find appropriate spinor variables changing the bar ones. We are not
going to apply such procedure to background fields which transformation we will find from T-dualization. In subsection 2.5 we will check
that both T-dual gravitinos satisfy single supersymmetry transformation rule.

%In the introductory considerations we explained the limited validity of our assumption of constant background fields. One of the conseqences is that supersymmetry invariants $d_\alpha$ and $\bar d_\alpha$ are not present in the 
%action and momenta $\pi_\alpha$ and $\bar\pi_\alpha$ survive and couple with the R-R field strength ${\rm P}^{\alpha\beta}$. Because supersymmetry invariants $d_\alpha$ and $\bar d_\alpha$ are variables in the most general case of pure spinor
%type II superstring action, we choose them to discuss transformation of bar variables. The same thing could be done using supersymmetry invariants $\Pi_\pm^\mu$ and commutation relations.

Note that according to \cite{berko,fTdual} fermionic coordinates, $\theta^\alpha$ and $\bar\theta^\alpha$, and their canonically conjugated momenta, $\pi_\alpha$ and $\bar \pi_\alpha$, are parts of supersymmetry
invariant variables
\begin{equation}\label{eq:susyd}
 d_\alpha =  \pi_\alpha -\frac{1}{2} (\Gamma_\mu \theta)_\alpha (\partial_+ x^\mu + \frac{1}{4} \theta \Gamma^\mu \partial_+ \theta ) \, \qquad
 \bar d_\alpha = \bar \pi_\alpha -\frac{1}{2} (\Gamma_\mu \bar \theta)_\alpha (\partial_- x^\mu + \frac{1}{4} \bar \theta \Gamma^\mu \partial_- \bar \theta )  \, .
\end{equation}

In T-dual theory, as a consequence of two types of $\Gamma$ matrices, there are two types supersymmetry invariant variables
\begin{equation}
{}_a d_\alpha={}_a \pi_\alpha-\frac{1}{2}({}_a\Gamma^{\hat\mu}\;{}_a\theta)_\alpha (\partial_+\;{}_a X_{\hat\mu}+\frac{1}{4}{}_a\theta {}_a\Gamma_{\hat\mu}\partial_+\;{}_a\theta)\, ,
\end{equation}
\begin{equation}
{}_a \bar d_\alpha={}_a \bar\pi_\alpha-\frac{1}{2}({}_a\bar\Gamma^{\hat\mu}\;{}_a\bar\theta)_\alpha (\partial_-\;{}_a X_{\hat\mu}+\frac{1}{4}{}_a\bar\theta {}_a\bar\Gamma_{\hat\mu}\partial_-\;{}_a\bar\theta)\, .
\end{equation}
We want to have both expressions with the same $\Gamma$ matrices. Using relation (\ref{eq:rgamam}) we can rewrite bar expressions as
\begin{equation}
({}_a \Omega \;{}_a \bar d)_\alpha = ({}_a \Omega\;{}_a\bar\pi)_\alpha-\frac{1}{2}({}_a\Gamma^{\hat\mu}{}_a\Omega\;{}_a\bar\theta)_\alpha(\partial_-\;{}_a X_{\hat\mu}+\frac{1}{4}{}_a\bar\theta\;{}_a \Omega^{-1}{}_a\Gamma_{\hat\mu}\;{}_a\Omega\;\partial_- {}_a\bar\theta)\, .
\end{equation}
So, if we preserve expressions for ${}_a\theta^\alpha=\theta^\alpha$ and ${}_a\pi_\alpha=\pi_\alpha$, change bar variables
\begin{equation}\label{eq:bulet1}
{}^\bullet \bar\theta^\alpha\equiv {}_a \Omega^\alpha{}_\beta \;{}_a\bar\theta^\beta\, ,\quad {}^\bullet \bar\pi_\alpha\equiv {}_a \Omega_\alpha{}^\beta\;{}_a \bar\pi_\beta\, ,
\end{equation}
and take
\begin{equation}\label{eq:norma1}
\Omega^2=1\, ,
\end{equation}
the transformation with bar variables will obtain the same form as those without bar in ${}_a d_\alpha$. Consequently, T-dual supersymmetric invariant variables ${}_a d_\alpha$ and ${}_a\Omega_\alpha{}^\beta\;{}_a\bar d_\beta$
are expressed in unique form in terms of true T-dual spinor variables $\theta^\alpha$, $\pi_\alpha$, ${}^\bullet \bar\theta^\alpha$ and ${}^\bullet \bar\pi_\alpha$
\begin{eqnarray}
{}_a d_\alpha=d_\alpha\, ,\quad {}^\bullet \bar d_\alpha={}_a \Omega_\alpha{}^\beta \bar d_\beta= {}^\bullet \bar\pi_\alpha -\frac{1}{2}({}_a \Gamma^{\hat\mu}\;{}^\bullet\bar\theta)_\alpha (\partial_-\;{}_a X_{\hat\mu}+\frac{1}{4}{}^\bullet\bar\theta\; {}_a \Gamma_{\hat\mu}\partial_-\;{}^\bullet\bar\theta)\, ,
\end{eqnarray}
if condition (\ref{eq:norma1}) is satisfied.

\subsubsection{Spinorial representation of the Lorentz transformation}

In order to find expressions for bar spinors in T-dual background we should first solve equation   (\ref{eq:omega}) and find expression for ${}_a \Omega$. We will do it
for $B_{\mu \nu} \to 0$, so that  $\hat \theta^{a b}_+ \to -\frac{1}{\kappa} (\widetilde{G}^{-1})^{ab}$, where $\widetilde{G}_{ab}$ is $ab$ component of
$G_{\mu \nu}$. Then from   (\ref{eq:lab}) it follows
\begin{equation}\label{eq:proj}
 {}_a \omega^{\underline{a}\underline{b}} \to e^{\underline{a}}{}_a  (\widetilde{G}^{-1})^{ab} (e^T)_b{}^{\underline{b}} \equiv {}_a P^{\underline{a} \underline{b}} \, ,
\end{equation}
where ${}_a P^{\underline{a}}{}_{ \underline{b}}$ is some $a$ dependent projector on the $\underline{a} \underline{b}$ subspace  ${}_a P^{\underline{a}}{}_{ \underline{c}} \,\, {}_a P^{\underline{c}}{}_{ \underline{b}} = {}_a P^{\underline{a}}{}_{ \underline{b}}$.
If we introduce $\Gamma$-matrices in curved space
\begin{eqnarray}\label{eq:obicnag}
\Gamma^\mu = (e^{-1})^\mu {}_{\underline{a}} \Gamma^{\underline{a}} \,   ,
\end{eqnarray}
%We can also introduce the projector ${}_i P^{\underline{a} \underline{b}} = \eta^{\underline{a} \underline{b}} - {}_a P^{\underline{a} \underline{b}}$, and the corresponding $\Gamma$-matrices
%\begin{equation}\label{eq:gagi}
%{}_a \Gamma^{\underline{a}} \equiv  {}_a P^{\underline{a} \underline{c}} \, \eta_{\underline{c} \underline{b}} \, \Gamma^{\underline{b}}  \, , \qquad
%{}_i \Gamma^{\underline{a}} \equiv  {}_i P^{\underline{a} \underline{c}} \, \eta_{\underline{c} \underline{b}} \, \Gamma^{\underline{b}} \, .
%\end{equation}
we can rewrite  expression  (\ref{eq:omega}) in the form
\begin{equation}\label{eq:omega0}
{}_a \Omega \, \Gamma^\mu  = \left[\Gamma^\mu - 2 \,((e^{-1})^\mu {}_{\underline{a}} \, {}_a P^{\underline{a}}{}_{\underline{b}}  \, \Gamma^{\underline{b}} \right]  \,   {}_a \Omega   \, .
\end{equation}
To simplify derivation from now on we will suppose that metric tensor is diagonal. Then $(e^{-1})^\mu {}_{\underline{a}} \, {}_a P^{\underline{a}}{}_{\underline{b}} = \delta^\mu_a (e^{-1})^a {}_{\underline{a}} $ and we have
\begin{equation}\label{eq:omega1}
{}_a \Omega \, \Gamma^\mu  = \left[\Gamma^\mu - 2 \,\delta^\mu_a \, \Gamma^a  \right]  \,   {}_a \Omega   \, .
\end{equation}

For $\mu =a$ and $\mu =i$ we obtain
\begin{equation}\label{eq:omega2}
{}_a \Omega \,\,  \Gamma^a  \, \,  = -  \Gamma^a  \,{}_a \Omega   \, ,   \qquad
{}_a \Omega \,\,  \Gamma^i   \, \,  =   \Gamma^i  \, \,{}_a \Omega   \, .
\end{equation}
%Let us now introduce the following gamma matrices
%\begin{equation}
%\Gamma_a\equiv e_a{}^{\underline a}\eta_{\underline{ab}}\;{}_a\Gamma^{\underline b}\, ,\quad \Gamma_i\equiv e_i{}^{\underline a}\eta_{\underline{ab}}{}_i \,\, \Gamma^{\underline b}\, ,
%\end{equation}
The  $\Gamma$-matrices in curved space for diagonal metric  satisfy the algebra
\begin{equation}
\{\Gamma^a,\Gamma^b\}=2(G^{-1})^{ab}\, ,\quad \{\Gamma^a,\Gamma^i\}=0\, ,\quad \{\Gamma^i,\Gamma^j\}=2 (G^{-1})^{ij}        \, .
\end{equation}

We should find such ${}_a \Omega$ that anticommutes with all matrices $\Gamma^a$ and commutes with all matrices $\Gamma^i$.
Let us first introduce  $\Gamma^{11}$ matrix as
\begin{equation}
\Gamma^{11}=(i)^{\frac{D(D-1)}{2}}\frac{1}{\prod_{\mu=0}^{D-1}G_{\mu\mu}} \varepsilon_{\mu_1\mu_2\dots \mu_D}\Gamma^{\mu_1}\Gamma^{\mu_2}\dots\Gamma^{\mu_D}\, ,
\end{equation}
where normalization constant is chosen so that $\Gamma^{11}$ satisfies the condition $(\Gamma^{11})^2=1$.

Then we define analogy of $\Gamma^{11}$ matrix in subspace spanned by T-dualized directions
\begin{equation}\label{eq:gama5}
{}_a  \Gamma \,  = (i)^{\frac{d(d-1)}{2}}  \prod_{i=1}^{d}  \Gamma^{a_i}  = (i)^{\frac{d(d-1)}{2}}\,  \Gamma^{a_1} \Gamma^{a_2} \cdots  \Gamma^{a_d} \,  ,
\end{equation}
so that
\begin{equation}\label{eq:gama52}
({}_a  \Gamma)^2 \,  =  \prod_{i=1}^{d} G^{a_i a_i} = \frac{1}{\prod_{i=1}^{d} G_{a_i a_i}} \, .
\end{equation}

Their commutation (anticommutation) relations with one $\Gamma$ matrix depend on number of coordinates $d$, along which we perform T-dualizations. Therefore we have
\begin{eqnarray}\label{eq:gama5cr}
& {}_a  \Gamma \,  \Gamma^a  = (-1)^{d+1}  \, \Gamma^a\, \, {}_a  \Gamma  \, ,   \qquad
& {}_a  \Gamma \,  \Gamma^i  = (-1)^{d}  \, \Gamma^i \, \, {}_a  \Gamma  \, ,     \,
\end{eqnarray}
which means that the solution of eq.(\ref{eq:omega2}) is proportional to
\begin{equation}\label{eq:solom0}
{}_a \Omega \,\, \sim  \, {}_a  \Gamma \,  ( \Gamma^{11})^d \, .
\end{equation}
Taking into account (\ref{eq:norma1}),  ${}_a \Omega^2 =1 $, we obtain
\begin{equation}\label{eq:solom}
{}_a \Omega \,\, = {\sqrt{\prod_{i=1}^{d} G_{a_i a_i} }} \,\, {}_a  \Gamma \,  (i \, \Gamma^{11})^d \, .
\end{equation}
This is a general solution. Note that for $a_1 \bigcap a_2 =0$ we have ${}_{a_1} \Omega  \,\, {}_{a_2} \Omega =(-1)^{d_1 d_2} {}_a  \Omega$, where
$a= a_1 \bigcup a_2$.

When the number of coordinates along which we perform T-duality is even $(d=2k)$,  we have ${}_a \Omega \,\, =  (-1)^{\frac{d}{2}}{\sqrt{\prod_{i=1}^{d} G_{a_i a_i} }}  \,\, {}_a  \Gamma$. As a consequence of the relation $\Gamma^{11} \, {}_a \Omega = (-1)^d \, \, {}_a \Omega \, \, \Gamma^{11}$ we can conclude that in that case bar spinors preserve chirality.
When the number of coordinates along which we perform T-duality is odd $(d=2k+1)$,  we have ${}_a \Omega \,\, = (-1)^{\frac{d-1}{2}}{\sqrt{\prod_{i=1}^{d} G_{a_i a_i} } } \, i\, \,  {}_a  \Gamma \,
\Gamma^{11}$. As a consequence of the above relation such transformation changes chirality of the bar spinors.

In the particular case, when we perform T-dualization along only one direction,
$x^{a_1}$, then $ {}_a  \Gamma \to \Gamma^{a_1}\,  , d \to 1$  and we obtain  the result well known in the literature \cite{S,H,BPT}
\begin{equation}\label{eq:solom1}
{}_{a_1} \Omega \,\, = i {\sqrt{G_{a_1 a_1} } }  \, \Gamma^{a_1}  \,  \Gamma^{11} \, .
\end{equation}
This is the case of the transition between IIA and IIB theory, when T-duality change chirality of the bar spinors.

When we perform T-dualization along all coordinates then $d\to D=10$, ${}_a  \Gamma   \to  \frac{\Gamma^{11}}{\sqrt{\prod_{\mu=0}^{D-1}G_{\mu\mu}}}$ and from (\ref{eq:solom}) we obtain
\begin{equation}\label{eq:solomg}
{}^\star \Omega \,\, =  (-1)^{\frac{D}{2}} \,\,  \Gamma^{11}=-\Gamma^{11} \,  \, .
\end{equation}

\subsection{Choice of the proper fermionic coordinates and T-dual background fields}

We have already learned that in order to have compatibility between supersymmetry and T-duality, we should choose the dual bar variables with "bullet" in accordance with (\ref{eq:bulet1}).
So, before we read the T-dual background fields,  we will reexpress the action   (\ref{eq:SBfixd}) in terms of the appropriate  spinor coordinates (\ref{eq:bulet1}) which, with the help of the relation ${}_a \Omega^2=1$,
produces
%\begin{equation}\label{eq:spcoo}
% {}_a^\bullet  \bar \pi_{\alpha}  \equiv  ({}_a \Omega^T)_\alpha{}^\beta \,\, {}_a  \bar \pi_{\beta}   \, , \qquad   {}_a^\bullet  \bar \theta^\alpha   \equiv  ( {}_a \Omega)^\alpha{}_\beta \,\, \bar \theta^\beta   \, ,
%\end{equation}
\begin{eqnarray}\label{eq:SBtd}
&{}& {}_a S  (y_a, x^i,\theta^\alpha,{}^\bullet  \bar \theta^\alpha,\pi_\alpha,{}^\bullet \bar \pi_\alpha) =   \nonumber \\
&{}& \int_\Sigma d^2\xi \left\{\frac{\kappa^2}{2}   \partial_+ y_a  \hat \theta_-^{ab}  \partial_- y_b   + \kappa^2 \partial_+ y_a \hat \theta_-^{ab}  \Pi_{+b j} \partial_- x^j   - \kappa^2 \partial_+ x^i \Pi_{+ia}
\hat \theta_-^{ab}  \partial_- y_b +\frac{1}{4\pi}\Phi R^{(2)}  \right.   \nonumber \\
&+&  \kappa \partial_+ x^i ( \Pi_{+ ij} -2\kappa  \Pi_{+ia}  \hat \theta_-^{ab} \Pi_{+b j}        )  \partial_- x^j   \nonumber \\
&-&  \pi_\alpha \partial_-(\theta^\alpha + \Psi^\alpha_i x^i  -2 \Psi^\alpha_a \hat \theta_-^{ab} \Pi_{+b j} x^j -  \Psi^\alpha_a  \hat \theta_-^{ab}  y_b   )        \nonumber \\
&+&  \partial_+ [{}^\bullet \bar \theta^{\gamma} {}_a\Omega_\gamma{}^\alpha  +\bar \Psi^{\alpha}_i x^i     + 2  \bar \Psi^\alpha_a \hat \theta_+^{ab} \Pi_{-b j} x^j + \bar \Psi^\alpha_a  \hat \theta_+^{ab}  y_b  ] {}_a \Omega_\alpha{}^\beta
{}^\bullet \bar \pi_{\beta}       \nonumber \\
&+&  \left. 2 \pi_\alpha \Psi^\alpha_a \hat \theta_-^{ab} \bar \Psi^\beta_b   {}_a \Omega_\beta {}^\gamma {}^\bullet \bar \pi_{\gamma}
       + \frac{1}{2\kappa} \, e^{\frac{\Phi}{2}} \, \pi_\alpha F^{\alpha \beta}    {}_a \Omega_\beta {}^\gamma {}^\bullet \bar \pi_{\gamma}  \right\}  \, \, .
\end{eqnarray}

Consequently, applying the Buscher T-dualization procedure \cite{B} along bosonic coordinates $x^a$ of the action (\ref{eq:SB}) the T-dual action obtains the form
\begin{eqnarray}\label{eq:TSB}
&{}&{}_a S= \int_\Sigma d^2\xi \left[ \kappa\partial_+  ({}_a X)_{\hat \mu}  \,{}_a \Pi^{\hat  \mu \hat \nu}_+\partial_- ({}_a X)_{\hat \nu}+\frac{1}{4\pi}{}_a \Phi R^{(2)}\right.
\\&-& \left. \pi_\alpha
\partial_- [\theta^\alpha+{}_a\Psi^{\alpha \hat \mu}
({}_a X)_{\hat \mu}]+\partial_+ [{}^\bullet \bar\theta^{\alpha}+ {}_a \bar \Psi^{\alpha \hat \mu}
({}_a X)_{\hat \mu}] \, {}^\bullet \bar \pi_{\alpha}+\frac{1}{2\kappa}   e^{\frac{{}_a \Phi}{2}}  \pi_\alpha \,\,  {}_a F^{\alpha \beta} \, {}^\bullet \bar \pi_{\beta}\right ]\, ,  \nonumber
\end{eqnarray}
where $({}_a X)_{\hat \mu}= (y_a, x^i)$, ${}_a \Psi^{\alpha \hat \mu}= ({}_a \Psi^{\alpha a}, {}_a \Psi^\alpha_i )$ and ${}_a \bar\Psi^{\alpha \hat \mu}= ({}_a  \bar\Psi^{\alpha a}, {}_a \bar\Psi^\alpha_i )$.

Now, we are ready to read the T-dual background fields
\begin{eqnarray}
&{}& {}_a \Pi_\pm^{ab}=\frac{\kappa}{2}\hat\theta_\mp^{ab}\, ,\label{eq:3}\\
&{}& {}_a \Pi_{\pm i}{}^a=-\kappa \Pi_{\pm ib} \hat\theta_\mp^{ba}\, ,\quad {}_a (\Pi_\pm)^a{}_i=\kappa\hat\theta_{\mp}^{ab}\Pi_{\pm bi}\, ,\label{eq:2}\\
&{}& {}_a \Pi_{\pm ij}=\Pi_{\pm ij}-2\kappa \Pi_{\pm ia}\hat\theta_{\mp}^{ab}\Pi_{\pm bj}\, ,\label{eq:1}\\
&{}& {}_a \Psi^{\alpha a}=\kappa \hat\theta^{ab}_+ \Psi^\alpha_b \, ,\quad {}_a {\bar\Psi}^{\alpha a}=\kappa \, {}_a \Omega^\alpha{}_\beta \, \hat\theta_-^{ab}
   \bar\Psi^{\beta}_b\, ,\label{eq:apartpsi}\\
&{}& {}_a \Psi^{\alpha}_i= \Psi^\alpha_i-2\kappa \Pi_{-ib}\hat\theta_+^{ba}\Psi^\alpha_a \, ,\quad
{}_a {\bar\Psi}^{\alpha}_i=  {}_a \Omega^\alpha{}_\beta ( \bar\Psi^\beta_i-2\kappa\Pi_{+ib}\hat\theta_-^{ba} \bar\Psi^\beta) \label{eq:partpsi}\, ,\\
&{}& e^{\frac{{}_a \Phi}{2}} {}_a F^{\alpha\beta}= (e^{\frac{\Phi}{2}} \,     F^{\alpha \gamma}+4\kappa \Psi^\alpha_a \hat\theta^{ab}_-\bar\Psi^\gamma_b ){}_a\Omega_\gamma{}^\beta \label{eq:dualFalfabeta}\, ,
\end{eqnarray}
when ${}_a \Omega$ is defined in (\ref{eq:solom}). 

The dilaton transformation in term $\Phi R^{(2)}$ originates from quantum theory and will be discussed in subsection 2.6.

\subsection{Supersymmetry transformations of T-dual gravitinos}

Note that in the expressions for T-dual fields ${}_a\bar\Psi^{\alpha a}$, ${}_a\bar\Psi^\alpha_i$ and ${}_a F^{\alpha\beta}$ the matrix ${}_a\Omega$ appears
as a consequence of T-dualization procedure and adoptions of "bullet" spinor coordinates. In Refs.\cite{H,BPT} it appears as a consequence of compatibility between supersymmetry and T-duality.

Supersymmetry transformation of gravitino is expressed in terms of covariant derivatives
\begin{equation}\label{eq:susy}
\delta_\varepsilon \Psi^\alpha_\mu = D_\mu \, \varepsilon^\alpha + \cdots  \, , \qquad  \delta_{\bar\varepsilon} \bar \Psi^\alpha_\mu = D_\mu \, \bar \varepsilon^\alpha + \cdots    \, ,
\end{equation}
with the same covariant derivative on both left and right  spinors
\begin{equation}\label{eq:code}
 D_\mu  = \partial_\mu \,   + \omega_\mu \,   \, .
\end{equation}

In the T-dual theory, as a consequence of two kinds of spin connections,  there are two kind of covariant derivatives
\begin{equation}
{}_a D^{\hat\mu}=\partial^{\hat\mu}+{}_a \omega^{\hat\mu}\, ,\quad {}_a \bar D^{\hat\mu}=\partial^{\hat\mu}+{}_a \bar\omega^{\hat\mu}\, ,
\end{equation}
such that
\begin{equation}
{}_a \delta_\varepsilon \;{}_a \Psi^{\alpha\hat\mu}={}_a D^{\hat\mu}\varepsilon^\alpha\, ,\quad {}_a\bar\delta_{\bar\varepsilon}\;{}_a\bar\Psi^{\alpha\hat\mu}={}_a \bar D^{\hat\mu}\bar\varepsilon^\alpha\, .
\end{equation}
Let us show that improvement with ${}_a\Omega$ in transformation of bar gravitionos just turns ${}_a\bar D^{\hat\mu}$ to ${}_a D^{\hat\mu}$.
In fact, from
\begin{equation}
{}_a \bar\delta_{\bar\varepsilon}\;{}_a\bar\Psi^{\alpha\hat\mu}={}_a\Omega^\alpha{}_\beta \left(\partial^{\hat\mu}\bar\varepsilon^\beta+{}_a\bar\omega^{\hat\mu\beta}{}_\gamma \bar\varepsilon^\gamma\right)\, ,
\end{equation}
with the help of (\ref{eq:oco}), for constant ${}_a\Omega$, we have
\begin{equation}
{}_a\bar\delta_{\bar\varepsilon}\;{}_a\bar\Psi^{\alpha\hat\mu}=\partial^{\hat\mu}({}_a\Omega^\alpha{}_\beta \bar\varepsilon^\beta)+{}_a\omega^{\hat\mu\alpha}{}_\beta \;{}_a\Omega^\beta{}_\gamma \bar\varepsilon^\gamma={}_a D^{\hat\mu}({}_a\Omega^\alpha{}_\beta \bar\varepsilon^\beta)={}_a\delta_{{}_a\Omega\bar\varepsilon}\;{}_a \bar\Psi^{\alpha\hat\mu}\, .
\end{equation}
Therefore, it is clear that in order to preserve the same spin connection for both chiralities we should additionally change bar supersymmetry
parameter
\begin{equation}\label{eq:barsp}
{}^\bullet  \bar \varepsilon^\alpha   \equiv   ({}_a \Omega)^\alpha{}_\beta \,\, {}_a\bar \varepsilon^\beta   \, .
\end{equation}

\subsection{Transformation of pure spinors}

In this subsection we will find transformation laws
for pure spinors, $\lambda^\alpha$ and $\bar\lambda^\alpha$, which are the main ingredient of pure spinor formalism.

It is well known that pure spinors satisfy so called pure spinor constraints
\begin{equation}\label{eq:psc}
\lambda^\alpha (\Gamma^\mu)_{\alpha\beta} \lambda^\beta=0\, ,\quad \bar\lambda^\alpha (\Gamma^\mu)_{\alpha\beta} \bar\lambda^\beta=0\, .
\end{equation}
After T-dualization they turn into
\begin{equation}\label{eq:psct}
{}_a\lambda^\alpha ({}_a \Gamma_{\hat\mu})_{\alpha\beta}\; {}_a \lambda^\beta=0\, ,\quad {}_a\bar\lambda^\alpha ({}_a \bar\Gamma_{\hat\mu})_{\alpha\beta}\; {}_a \bar\lambda^\beta=0\, .
\end{equation}
The relation between matrices ${}_a\Gamma_{\hat\mu}$ and ${}_a\bar\Gamma_{\hat\mu}$ is given in (\ref{eq:rgamam}).
In order to have both expressions with same gamma matrices, as before, we preserve the expression for unbar variables
\begin{equation}
{}_a\lambda^\alpha=\lambda^\alpha\, .
\end{equation}
and change bar variables
\begin{equation}\label{eq:barlambda}
{}^\bullet \bar\lambda^\alpha={}_a\Omega^\alpha{}_\beta \;{}_a\bar\lambda^\beta\, .
\end{equation}

The varables $w_\alpha$ and $\bar w_\alpha$ are canonically conjugated momenta to the pure spinors $\lambda^\alpha$ and $\bar\lambda^\alpha$, respectively. 
The transformation laws for pure spinor momenta can be found from the expressions for $N_+^{\mu\nu}$ and $\bar N_-^{\mu\nu}$ (\ref{eq:Npm})
which would appear in the action if we would take higher power terms in $\theta^\alpha$ and $\bar\theta^\alpha$. After T-dualization these expressions become
\begin{equation}
{}_a N_{+\hat\mu\hat\nu}=\frac{1}{2}{}_a w_\alpha ({}_a \Gamma_{[\hat\mu\hat\nu]})^\alpha{}_\beta \;{}_a\lambda^\beta\, ,\quad {}_a \bar N_{-\hat\mu\hat\nu}=\frac{1}{2}{}_a \bar w_\alpha ({}_a \bar\Gamma_{[\hat\mu\hat\nu]})^\alpha{}_\beta \;{}_a\bar\lambda^\beta\, .
\end{equation}
Using Eq.(\ref{eq:rgamam}) and definition of $\Gamma^{[\mu\nu]}$ (\ref{eq:bazis}) we see that relation between ${}_a \Gamma_{[\hat\mu\hat\nu]}$ and ${}_a \bar\Gamma_{[\hat\mu\hat\nu]}$ is the same as between 
gamma matrices (\ref{eq:rgamam}).
As in the previous case, in order to have unique set of gamma matrices, we do not change unbar variables
\begin{equation}
{}_a w_\alpha=w_\alpha\, .
\end{equation}
while choose bar variables in the form
\begin{equation}
{}^\bullet \bar w_\alpha={}_a \Omega_\alpha{}^\beta \;{}_a\bar w_\beta\, .
\end{equation}

Let us note that free field actions $S_\lambda$ and $S_{\bar\lambda}$ are invariant under T-dualization because ${}_a\Omega^2=1$.

\subsection{T-dual transformation of antisymmetric fields \\-- from IIB to IIA theory}

To find the T-dual transformation laws for antisymmetric fields we will start with expression (\ref{eq:dualFalfabeta}).
First, as it is explained in Refs.\cite{B,GR} the quantization procedure produces the well known  shift in the dilaton transformation
\begin{equation}\label{eq:dsh}
 {}_a \Phi  = \Phi - \ln \det ( 2 \Pi_{+ ab}) = \Phi -  \ln \sqrt{\frac{\det G_{a b}}{\det {}_a G^{a b}}} \, .
\end{equation}
Together with (\ref{eq:dualFalfabeta}) it gives relation between initial and T-dual background fields
\begin{eqnarray}\label{eq:FG}
&{}&     {}_a F^{\alpha\beta}=    \,   \,  \sqrt[4]{\frac{ \det G_{a b}}{\det {}_a G^{a b}} }  \, \, (   F^{\alpha\gamma}+4e^{-\frac{\Phi}{2}}\kappa \Psi^\alpha_a \hat\theta^{ab}_-\bar\Psi^\gamma_b ){}_a\Omega_\gamma{}^\beta \, .
\end{eqnarray}
For $B_{\mu\nu}=0$ we have ${}_a G^{a b}= (G_E^{-1})^{a b} = (G^{-1})^{a b}$, and consequently $ \sqrt[4]{\frac{ \det G_{a b}}{\det {}_a G^{a b}} }= \sqrt[4]{ (\det G_{a b})^2} = \sqrt{\mid \det G_{a b}   \mid }$.
It is important to stress that unlike in expression (\ref{eq:solom}) for ${}_a \Omega$ here we have absolute value under square root. For  diagonal metric $G_{\mu\nu}$ we have
$\det G_{a b} = \prod_{i=1}^d G_{a_i a_i} $ and taking into account expression (\ref{eq:solom}) we find
\begin{equation}\label{eq:tdf}
{}_a F^{\alpha\beta}=i^d\sqrt{sign ({\prod_{i=1}^{d} G_{a_i a_i} })} \,\, {\prod_{i=1}^{d} G_{a_i a_i} } \left(F^{\alpha\gamma}+4e^{-\frac{\Phi}{2}}\kappa\Psi^\alpha_a \hat\theta^{ab}_-\bar\Psi^\gamma_b\right)   \,(  {}_a  \Gamma \,  \Gamma_{11}^d )_\gamma{}^\beta\, .
\end{equation}

Note that we are going to T-dualize all $D$-directions. Then it is necessary to perform T-dualization along time-like direction. Here the above square root has important consequences.
For our signature  $(+,-,-,\dots,-)$,   the square of the field strength $({}_a F^{\alpha\beta})^2$ and, consequently the square of
all antisymmetric fields will change the sign when we perform T-dualization along time-like direction. This is just what we need to obtain Type $II^\star$ theories in accordance with Ref.\cite{timelike}.

In  a simple case when gravitino fields and  Kalb-Ramond field are zero and metric is diagonal we will express transition from type IIB to type IIA theory.
Taking $d=1$ we have
\begin{equation}\label{eq:tdf0}
{}_a F^{\alpha\beta}=i\sqrt{sign ( G_{aa} )} \,\,  G_{aa}  \,\,  F^{\alpha\gamma}      \,(\Gamma^{11}\Gamma^a)_\gamma{}^\beta\, .
\end{equation}

Let us choose type IIB as a starting theory. The matrix $\Gamma^{11}$ turns $F^{(n)}$ to $F^{(10-n)}$ where
\begin{equation}\label{eq:asfB}
(\textsc{F}^{(n)})^{\alpha\beta} =  \frac{1}{n!}  F_{\mu_1 \mu_2 \cdots \mu_n} (\Gamma^{[\mu_1 \mu_2 \cdots \mu_n}])^{\alpha\beta}   \, .
\end{equation}
As a consequence of the chirality condition $F\Gamma^{11}=-\Gamma^{11}F$ the independent tensors are $F^{(1)}$, $F^{(3)}$ and self dual part of $F^{(5)}$.
So we can write
\begin{equation}\label{eq:IIB}
F^{\alpha\gamma}(\Gamma^{11})_\gamma{}^\beta =   \left( \textsc{F}^{(1)} + \textsc{F}^{(3)} +\frac{1}{2} \textsc{F}^{(5)} \right)^{\alpha\beta} \, ,
\end{equation}
Similarly, in T-dual theory  (here it is IIA)  we have
\begin{equation}\label{eq:IIA}
{}_a F^{\alpha\beta} = \left({}_a \textsc{F}^{(2)} +  {}_a \textsc{F}^{(4)}  \right)^{\alpha\beta} \, ,
\end{equation}
where now
\begin{equation}\label{eq:asfA}
 ({}_a \textsc{F}^{(n)})^{\alpha\beta} =  \frac{1}{n!} {}_a F^{\hat \mu_1 \hat \mu_2 \cdots \hat \mu_n} ({}_a \Gamma_{[\hat \mu_1 \hat \mu_2 \cdots \hat \mu_n}])^{\alpha\beta}   \, .
\end{equation}

The $\Gamma$-matrices on both sides are defined in curved space. For initial theory it is just (\ref{eq:obicnag})
%$\Gamma^\mu = (e^T)^{\mu} {}_{ \underline{a}} \Gamma^{\underline{a}}$,
while for T-dual theory  it is defined in the first relation  (\ref{eq:gamam1}) as ${}_a \Gamma_{\hat \mu} =  ({}_a e^{-1})_{\hat \mu \underline{a}} \, \Gamma^{\underline{a}}$.
As a consequence of the first relation (\ref{eq:tde}) between vielbeins ${}_a e^{\underline{a} \hat \mu} = e^{\underline{a}}{}_\nu (Q^T)^{\nu \hat \mu}$  we can find the relation between $\Gamma$-matrices
\begin{equation}\label{eq:rgamm}
{}_a \Gamma_{\hat \mu}    =   (Q^{-1 T})_{\hat \mu \nu} \Gamma^\nu   \, ,
\end{equation}
which  produces
\begin{equation}\label{eq:asfA1}
 ({}_a \textsc{F}^{(n)})^{\alpha\beta} =  \frac{1}{n!} ({}_a^Q F)_{ \mu_1  \mu_2 \cdots  \mu_n} ( \Gamma^{[\mu_1  \mu_2 \cdots  \mu_n]})^{\alpha\beta}   \, ,
\end{equation}
where
\begin{equation}\label{eq:FQF}
({}_a^Q F)_{ \mu_1  \mu_2 \cdots  \mu_n}  = {}_a F^{\hat \mu_1 \hat \mu_2 \cdots \hat \mu_n}   (Q^{-1 T})_{\hat \mu_1 \mu_1} (Q^{-1 T})_{\hat \mu_2 \mu_2} \cdots (Q^{-1 T})_{\hat\mu_n \mu_n}  \, .
\end{equation}

Using the standard relation between $\Gamma$-matrices
\begin{equation}\label{eq:rGm}
\Gamma^{[\mu_1 \mu_2 \cdots \mu_n]}\Gamma^a  = \Gamma^{\mu \mu_1 \mu_2 \cdots \mu_n a} - \frac{1}{(n-1)!} G^{a [ \mu_n} \Gamma^{\mu_1 \mu_2 \cdots \mu_{n-1} ]} \, ,
\end{equation}
we obtain
\begin{equation}\label{eq:kraj}
F^{(n)}\Gamma^a=\frac{1}{n!}F_{\mu_1\mu_2\cdots \mu_n}\Gamma^{[\mu_1\mu_2\cdots\mu_n a]}-\frac{1}{(n-1)!}F_{\mu_1\mu_2\cdots\mu_{n-1}}{}^a \;\Gamma^{[\mu_1\mu_2\cdots\mu_{n-1}]}\, .
\end{equation}

So, from (\ref{eq:tdf0}), (\ref{eq:IIB}), (\ref{eq:IIA}), (\ref{eq:asfA1}), (\ref{eq:FQF}) and (\ref{eq:kraj}) we can find general relation connected antisymmetric fields of Type IIA and type IIB theories
\begin{equation}\label{eq:fanf}
 {}_a \textsc{F}^{\hat \mu_1 \hat \mu_2 \cdots \hat \mu_n}    =  \sqrt{sign G_{aa}}G_{aa}
  \left( nF_{\mu_1\mu_2\cdots\mu_{n-1}} \delta^a{}_{\mu_n} - F_{\mu_1\mu_2\cdots\mu_n}{}^a \right) (Q^T)^{\mu_1 \hat \mu_1} (Q^T)^{\mu_2 \hat \mu_2} \cdots (Q^T)^{\mu_n \hat \mu_n}  \, .
\end{equation}

Under our assumptions we have
\begin{equation}
(Q^T)^{\mu\hat\mu}=\left(
\begin{array}{cc}
-G^{aa} & 0\\
0 & \delta_i{}^j
\end{array}\right)\, ,
\end{equation}
and consequently
\begin{equation}
{}_a F_{ij}=-i\sqrt{sign G_{aa}}G_{aa}F_{ij}{}^a\, ,\quad {}_a F_i{}^a=-2i\sqrt{sign G_{aa}} F_i\, ,
\end{equation}
\begin{equation}
{}_a F_{ijkq}=-\frac{i}{2}\sqrt{sign G_{aa}}G_{aa}F_{ijkq}{}^a\, ,\quad {}_a F_{ijk}{}^a=-4i\sqrt{sign G_{aa}} F_{ijk}\, .
\end{equation}
For the space-like directions $G_{aa} < 0$ and  $i \sqrt{sign G_{aa}}$ is real. For time-like direction $\sqrt{sign G_{aa}}\to \sqrt{sign G_{00}}=1$  and  remaining imaginary unit 
causes that squares of the antisymmetric fields get additional minus sign and type II theories swap to type $II^\star$ ones \cite{timelike}.

%with signature $(+,-,-,\dots,-)$

%%%%%%%%%%%%%%%%%%%%%%%%%%%%%%%%%%%%%%%%%%%%%%%%%%%%%%%%%%%%%%%%%%%%%%%%%%%%%%
%\cleq

\section{Double space formulation}
\setcounter{equation}{0}

In this section we will introduce double space, doubling all bosonic coordinates $x^\mu$ by corresponding T-dual ones $y_\mu$. We will rewrite the transformation laws  in double space and
show that both the equations of motion and Bianchi identities can be written by that single equation.

\subsection{T-dualization along all bosonic directions}

Applying the Buscher T-dualization procedure \cite{B} along all bosonic coordinates of the action (\ref{eq:SB}) the T-dual action
%\begin{eqnarray}\label{eq:TSB}
%&{}&{}^\star S=\kappa \int_\Sigma d^2\xi \partial_+ y_\mu \,{}^\star\Pi^{\mu\nu}_+\partial_- y_\nu
%\\&+&\int_\Sigma d^2 \xi \left[ -\pi_\alpha
%\partial_-(\theta^\alpha+{}^\star\Psi^{\alpha\mu}
%y_\mu)+\partial_+(\bar\theta^{\alpha}+{}^\star\bar \Psi^{\alpha\mu}
%y_\mu)\bar\pi_{\alpha}+\frac{1}{2\kappa}\pi_\alpha {}^\star F^{\alpha
%\beta}\bar \pi_{\beta}\right ]\, ,\nonumber
%\end{eqnarray}
has been obtained in Ref.\cite{nasnpb}. This is particular case of our relations (\ref{eq:3})-(\ref{eq:dualFalfabeta})  where T-dual background fields are of the form
\begin{equation}\label{eq:starpi}
{}^\star \Pi^{\mu\nu}_\pm\equiv {}^\star B^{\mu\nu}\pm \frac{1}{2}{}^\star G^{\mu\nu}=\frac{\kappa}{2}\Theta^{\mu\nu}_\mp\, ,
\end{equation}
\begin{equation}\label{eq:psibarpsidualf}
{}^\star\Psi^{\alpha\mu}=\kappa \Theta^{\mu\nu}_+ \Psi^\alpha_\nu\, ,\quad {}^\star \bar\Psi^{\alpha\mu}=\kappa \,\, {}^\star \Omega^\alpha {}_\beta  \, \Theta^{\mu\nu}_- \bar\Psi^\beta_\nu\, ,
\end{equation}
\begin{equation}\label{eq:Falfabeta}
e^{\frac{{}^\star \Phi}{2}} \, {}^\star F^{\alpha\beta}= (e^{\frac{\Phi}{2}}  F^{\alpha\gamma}+4\kappa \Psi^\alpha_\mu \Theta^{\mu\nu}_- \bar\Psi^\gamma_\nu) {}^\star \Omega_\gamma {}^\beta \, .
\end{equation}
Here we use the notation
\begin{equation}\label{eq:efpp}
G^E_{\mu\nu}=G_{\mu\nu}-4(BG^{-1}B)_{\mu\nu}\, ,\quad \Theta^{\mu\nu}=-\frac{2}{\kappa}(G_E^{-1}BG^{-1})^{\mu\nu}\, , \quad  {}^\star \Omega \,\, =  - \Gamma^{11} \,  ,
\end{equation}
and
\begin{equation}\label{eq:tetapm}
\Theta^{\mu\nu}_\pm=-\frac{2}{\kappa}(G_E^{-1}\Pi_{\pm}G^{-1})^{\mu\nu}=\Theta^{\mu\nu}\mp\frac{1}{\kappa}(G_E^{-1})^{\mu\nu}\, ,
\end{equation}
so that
\begin{equation}
(\Pi_{\pm}\Theta_{\mp})_\mu{}^\nu=\frac{1}{2\kappa}\delta_\mu{}^\nu\, .
\end{equation}
From (\ref{eq:starpi}) and (\ref{eq:tetapm}) it follows
\begin{equation}\label{eq:GBdualf}
{}^\star G^{\mu\nu}=(G_E^{-1})^{\mu\nu}\, ,\quad {}^\star B^{\mu\nu}=\frac{\kappa}{2}\Theta^{\mu\nu}\, .
\end{equation}

In this case the transformation laws  (\ref{eq:lawsb1}) and (\ref{eq:lawsb2})  (the relations between initial $x^\mu$ and T-dual coordinates $y_\mu$) obtain the form
\begin{eqnarray}\label{eq:xtdual}
\partial_{\pm}x^\mu \cong
-\kappa\Theta^{\mu\nu}_{\pm}  \partial_{\pm} y_\nu+\kappa \Theta^{\mu\nu}_\pm J_{\pm \nu}\, ,  \qquad  \partial_{\pm}y_\mu\cong
-2\Pi_{\mp\mu\nu} \partial_{\pm} x^\nu+J_{\pm \mu} \,  .
\end{eqnarray}

\subsection{Transformation laws in double space}

Rewriting equations (\ref{eq:xtdual}) in the form where terms multiplied by $\varepsilon_\pm{}^\pm=\pm 1$ are on the left-hand side of the equation, we obtain
\begin{equation}\label{eq:tdualc1}
\pm\partial_\pm y_\mu\cong G_{E \mu\nu}\partial_\pm x^\nu-2(BG^{-1})_\mu{}^\nu \partial_\pm y_\nu +2(\Pi_{\pm}G^{-1})_\mu{}^\nu J_{\pm \nu}\, ,
\end{equation}
\begin{equation}\label{eq:tdualc2}
\pm \partial_{\pm} x^\mu\cong (G^{-1})^{\mu\nu}\partial_\pm y_\nu+2(G^{-1}B)^\mu{}_\nu \partial_\pm x^\nu-(G^{-1})^{\mu\nu}J_{\pm \nu}\, .
\end{equation}

Let us introduce double space coordinates
\begin{equation}\label{eq:escoor}
Z^M=\left (
\begin{array}{c}
 x^\mu  \\
y_\mu
\end{array}\right )\, ,
\end{equation}
which contain all initial and T-dual coordinates.
In terms of double coordinates the relations (\ref{eq:tdualc1}) and (\ref{eq:tdualc2}) are replaced by one
\begin{equation}\label{eq:tdual}
\partial_\pm Z^M\cong \pm \Omega^{MN}\left({\cal H}_{NP} \partial_\pm Z^P+J_{\pm N}\right)\, ,
\end{equation}
where the matrix ${\cal H}_{MN}$ is known in literatute as generalized metric and has the form
\begin{equation}\label{eq:gm}
{\cal{H}}_{MN}  = \left (
\begin{array}{cc}
 G^E_{\mu \nu}  &  -2 \,  B_{\mu\rho}  (G^{-1})^{\rho \nu}  \\
2 (G^{-1})^{\mu \rho} \,  B_{\rho \nu}  & (G^{-1})^{\mu \nu}
\end{array}\right )\, .
\end{equation}
The double current $J_{\pm M}$ is defined as
\begin{equation}\label{eq:currentJ}
J_{\pm M}=\left(
\begin{array}{c}
2(\Pi_\pm G^{-1})_\mu{}^\nu J_{\pm \nu}\\
-(G^{-1})^{\mu\nu}J_{\pm \nu}
\end{array}
\right)\, ,
\end{equation}
and
\begin{equation}
\Omega^{MN}= \left (
\begin{array}{cc}
0 &  1_D \\
1_D  & 0
\end{array}\right )\, ,
\end{equation}
is constant symmetric matrix. Here $1_D$ denotes the identity operator in $D$ dimensions. 
Let us stress that matrix ${}_a \Omega$ and $\Omega^{MN}$ are different quantities.

By straightforward calculation we can prove the relations
\begin{equation}\label{eq:sodd}
{\cal{H}}^T  \Omega {\cal{H}} =\Omega \, ,\quad \Omega^2=1\, ,\quad \det {{\cal H}_{MN}}=1\, ,
\end{equation}
which means that ${\cal H}\in SO(D,D)$. In calculation of determinant we use the rule for block matrices
\begin{equation}
\det \left (
\begin{array}{cc}
A & B\\
C & D
\end{array}\right)=\det D \det(A-BD^{-1}C)\, .
\end{equation}
In Double Field Theory $\Omega^{MN}$ is the $SO(D,D)$ invariant metric and denoted by $\eta^{MN}$.

%%%%%%%%%%%%%%%%%%%%%%%%%%%%%%%%%%%%%%%%%%%%%%%%%%%%%%%%%%%%%%%%%%%%%%%%%%%%%%

\subsection{Equations of motion and double space action}

It is well known that equations of motion of initial theory are Bianchi identities in T-dual picture and vice versa \cite{ALLP,DS1,Duff,GR}.
As a consequence of the identity
\begin{equation}
\partial_+ \partial_- Z^M-\partial_-\partial_+ Z^M=0\, ,
\end{equation}
known as Bianchi identity, and relation (\ref{eq:tdual}), we obtain the consistency condition
\begin{equation}\label{eq:tdual1}
\partial_+\left[{\cal H}_{MN}\partial_- Z^N+J_{-M}\right]+\partial_-\left[{\cal H}_{MN}\partial_+ Z^N+J_{+M}\right]=0\, .
\end{equation}
In components it takes a form
\begin{eqnarray}\label{eq:em}
&{}&\partial_+  \partial_- x^\mu =-\frac{1}{\kappa}(G^{-1})^{\mu\nu}\left(\bar\Psi^\alpha_\nu\partial_+ \bar \pi_\alpha +\Psi^\alpha_\mu\partial_- \pi_\alpha\right)  \,  , \nonumber \\ &{}&  \partial_+   \partial_- y_\nu =-\frac{1}{\kappa}G^E_{\mu\nu}\left({}^\star \bar\Psi^{\alpha\mu}\partial_+\bar \pi_\alpha+{}^\star\Psi^{\alpha\mu}\partial_- \pi_\alpha\right) \,  .
\end{eqnarray}
These equations are equations of motion of the initial and T-dual theory. Double space formalism enables us to write both equations of motion and Bianchi identities by single relation (\ref{eq:tdual}).

The equation (\ref{eq:tdual1}) is equation of motion of the following action
\begin{equation}\label{eq:exact}
S=\frac{\kappa}{4}\int d^2\xi \left[\partial_+ Z^M {\cal H}_{MN}\partial_-Z^N+\partial_+ Z^M J_{-M}+J_{+M}\partial_- Z^M+{L}(\pi_\alpha,\bar \pi_\alpha)\right]\, ,
\end{equation}
where ${L}(\pi_\alpha,\bar \pi_\alpha)$ is arbitrary functional of fermionic momenta.

%%%%%%%%%%%%%%%%%%%%%%%%%%%%%%%%%%%%%%%%%%%%%%%%%%%%%%%%%%%%%%%%%%%%%%%%%%%%%%

\section{T-dualization of type II superstring theory as permutation of coordinates in double space}
\setcounter{equation}{0}

In this section we will derive the transformations of the generalized metric and current, which are consequence of the
permutation of some subset of the bosonic coordinates with the corresponding T-dual ones.
First we will present the method in the case of the complete T-dualization
(along all bosonic coordinates) and find the expressions for T-dual background fields.
Then we will apply the receipt on the case of partial T-dualization.

\subsection{The case of complete T-dualization}

In order to exchange all initial and T-dual coordinates let us introduce the permutation matrix
\begin{equation}
{\cal T}^M{}_N=\left(
\begin{array}{cc}
0 & 1_D\\
1_D & 0
\end{array}\right)\, ,
\end{equation}
so that double T-dual coordinate ${}^\star Z^M$ is obtained as
\begin{equation}\label{eq:simetry1}
{}^\star Z^M = {\cal T}^M{}_N Z^N=\left(
\begin{array}{c}
y_\mu\\
x^\mu
\end{array}\right) .
\end{equation}
We demand that T-dual transformation law for double T-dual coordinate ${}^\star Z^M$ has the same form as for initial coordinate $Z^M$ (\ref{eq:tdual})
\begin{equation}
\partial_\pm {}^\star Z^M\cong \pm \Omega^{MN} \left({}^\star {\cal H}_{NP} \partial_\pm {}^\star Z^P+{}^\star J_{\pm N}\right)\, .
\end{equation}
Then the T-dual generalized metric ${}^\star {\cal H}_{MN}$ and T-dual current ${}^\star J_{\pm M}$ are
\begin{equation}\label{eq:simetry2}
{}^\star {\cal H}_{MN}={\cal T}_M{}^K {\cal H}_{KL}{\cal T}^L{}_N\, ,\quad {}^\star J_{\pm M}={\cal T}_M{}^N J_{\pm N}\, .
\end{equation}
Permutation of the coordinates (\ref{eq:simetry1}) together with transformations of the background fields (\ref{eq:simetry2}) represents the symmetry transformations of the action (\ref{eq:exact}).

Using the corresponding expressions for ${\cal T}^M{}_N$, ${\cal H}_{MN}$ and $J_{\pm M}$, we obtain from the generalized metric transformation
\begin{equation}
{}^\star G^{\mu\nu}=(G_E^{-1})^{\mu\nu}\, ,\quad {}^\star B^{\mu\nu}=\frac{\kappa}{2}\Theta^{\mu\nu}\, .
\end{equation}
Taking into account that as a consequence of (\ref{eq:bulet1}) the bar dual variable is ${}^{\bullet \star} \bar \pi_\alpha = ({}^\star \Omega^T)_\alpha {}^\beta \bar \pi_\beta $,
from current transformations we have
\begin{equation}
{}^\star \Psi^{\alpha\mu}=\kappa \Theta^{\mu\nu}_+ \Psi^\alpha_\nu\, ,\quad {}^\star \bar\Psi^{\alpha\mu}=\kappa \, \, {}^\star \Omega^\alpha {}_\beta  \Theta^{\mu\nu}_- \bar\Psi^\beta_\nu\, ,
\end{equation}
where ${}^\star \Omega =-\Gamma^{11}$.

Consequently, using double space we can easily reproduce the results of T-dualization (\ref{eq:GBdualf})  and (\ref{eq:psibarpsidualf}). The problem with T-dualization of the R-R field strength $F^{\alpha\beta}$ will be disscussed in subsection 5.3.

\subsection{The case of partial T-dualization}

Applying the procedure presented in the previous subsection to the arbitrary subset of bosonic coordinates we will, in fact, describe all possible bosonic T-dualizations.
Let us split  coordinate index $\mu$ into $a$ and $i$  ( $a=0,\cdots,d-1$,  $i=d,\cdots,D-1$)
and denote T-dualization along direction $x^a$ and $y_a$ as
\begin{equation}\label{eq:circ}
{\cal T}^{a}=T^a \circ T_a  \,  ,  \quad    T^a \equiv T^{0}\circ T^{1}\circ\cdots\circ T^{d-1} \,  ,  \quad
T_a \equiv T_{0}\circ T_{1}\circ\cdots\circ T_{d-1} \,  ,
\end{equation}
where $\circ$ marks the operation of composition of T-dualizations.
Permutation of the initial coordinates $x^a$ with its T-dual $y_a$ we realize by multiplying double space coordinate
by the constant symmetric matrix $({\cal T}^a)^M{}_N$
\begin{equation}\label{eq:taua}
{}_a Z^M\equiv \left (
\begin{array}{c}
 y_a  \\
 x^i  \\
x^a   \\
y_i
\end{array}\right ) = ({\cal T}^a)^M{}_N Z^N\equiv \left (
\begin{array}{cccc}
0  & 0  & 1_a & 0  \\
0  & 1_i  & 0 & 0  \\
1_a & 0  & 0 & 0  \\
0  & 0  & 0 & 1_i
\end{array}\right )\left (
\begin{array}{c}
 x^a  \\
 x^i  \\
y_a   \\
y_i
\end{array}\right )\,  ,
\end{equation}
where $1_a$ and $1_i$ are identity operators in the subspaces spanned by $x^a$ and $x^i$, respectively. It is easily to check the following relations
\begin{equation}\label{eq:tprop}
({\cal T}^{a}  {\cal T}^{a})^M {}_N  = \delta^M {}_N  \, , \qquad \quad   ( {\cal T}^{a} \Omega {\cal T}^{a})^M {}_N  = \Omega^M {}_N     \, .
\end{equation}
The first relation means that after two T-dualizations we get the initial theory, while the second relation means that ${\cal T}^{a} \in O(D,D)$.

Let us apply the same approach as in the case of the full T-dualization presented in the previous subsection. We demand that double T-dual coordinate ${}_a Z^M$
satisfy the T-duality transformations of the form as initial one $Z^M$ (\ref{eq:tdual})
\begin{equation}\label{eq:tdualdf}
\partial_{\pm}\; {}_a Z^M \cong \pm \, \Omega^{MN} \left({}_a {\cal{H}}_{NK} \,\partial_{\pm}\;{}_aZ^K + {}_a J_{\pm N}\right)\, .
\end{equation}
Consequently, we find the T-dual generalized metric
\begin{equation}\label{eq:dualgm}
{}_a {\cal{H}}_{MN} =  ({\cal T}^{a})_M{}^K  {\cal{H}}_{KL} ({\cal T}^{a})^L{}_N \, ,
\end{equation}
and T-dual current
\begin{equation}\label{eq:parTJ}
{}_a J_{\pm M}=({\cal T}^a)_M{}^N J_{\pm N}\, .
\end{equation}
Note that equations (\ref{eq:taua}), (\ref{eq:dualgm}) and (\ref{eq:parTJ}) are symmetry transformations of the action (\ref{eq:exact}). The left subscript $a$ means dualization along $x^a$ directions.

%%%%%%%%%%%%%%%%%%%%%%%%%%%%%%%%%%%%%%%%%%%%%%%%%%%%%%%%%%%%%%%%%%%%%%%%%%%%%%

\section{T-dual background fields}
\setcounter{equation}{0}

In this section we will show that permutation of some bosonic coordinates leads to the same T-dual background fields as standard Buscher procedure \cite{nasnpb}.
The transformation of the generalized metric (\ref{eq:dualgm}) produces expressions for NS-NS T-dual background fields ($G_{\mu\nu}$ and $B_{\mu\nu}$). They are the
same as in bosonic string case obtained in Ref.\cite{sazdam}.
So, we will just shortly repeat these results.
From the transformation of the current $J_{\pm M}$ (\ref{eq:parTJ}) we will find T-dual background fields of the NS-R sector ($\Psi^\alpha_\mu$ and $\bar\Psi^\alpha_\mu$).
Because R-R field strength $F^{\alpha\beta}$ does not appear in T-dual transformations, we will find its T-dual under some assumptions.

\subsection{T-dual NS-NS background fields $G_{\mu\nu}$, $B_{\mu\nu}$}

Demanding that the T-dual generalized metric ${}_a {\cal H}_{MN}$ has the same form as the initial one ${\cal H}_{MN}$ (\ref{eq:gm}) but in terms of the T-dual fields
\begin{equation}\label{eq:h0d}
{}_a {\cal{H}}_{MN} = \left (
\begin{array}{cc}
{}_a G_E^{\mu \nu}  &  -2  ({}_a B \,{}_a G^{-1})^\mu{}_\nu   \\
2 ({}_a G^{-1}\, {}_a B)_\mu {}^\nu  & ({}_a G^{-1})_{\mu \nu}
\end{array}\right )\, ,
\end{equation}
and using Eq.(\ref{eq:dualgm}), one finds expressions for the NS-NS T-dual background fields ${}_a \Pi^{\mu\nu}_\pm$ in terms of the initial ones
\begin{equation}\label{eq:pipm}
{}_a \Pi_{ \pm}^{\mu \nu}  = \left (
\begin{array}{cc}
 {\tilde g}^{-1} \beta_1 D^{-1} \gamma - A^{-1} ({\tilde \beta} \mp \frac{1}{2})         &   \frac{1}{2} A^{-1} g^T -2  {\tilde g}^{-1} \beta_1 D^{-1} ({\bar \beta}^T \mp \frac{1}{2})    \\
 \frac{1}{2}D^{-1} \gamma -2 \bar{\gamma}^{-1} \beta_1^T A^{-1} ({\tilde \beta} \mp \frac{1}{2})      &   \bar{\gamma}^{-1} \beta_1^T A^{-1}g^T - D^{-1} ({\bar \beta}^T \mp \frac{1}{2})
\end{array}\right )\, ,
\end{equation}
where $\gamma$ and $\bar\gamma$ are defined in (\ref{eq:gnmj}), $g$ and $\tilde g$ in (\ref{eq:gdef}),
while $\beta_1$, $\tilde\beta$ and $\bar\beta$ are defined in (\ref{eq:bgnmj}).
The quantities $A$ and $D$ are given in (\ref{eq:velA}) and (\ref{eq:DD}), respectively.
In more compact form we have
\begin{equation}\label{eq:pipmr}
{}_a \Pi_{ \pm}^{\mu \nu}  = \left (
\begin{array}{cc}
 \frac{\kappa}{2} {\hat \theta}_{ \mp}^{ab}   &  \kappa {\hat \theta}_{\mp}^{ab} \Pi_{\pm bi}  \\
-\kappa \Pi_{\pm ib} {\hat \theta}_{ \mp}^{ba}    &   \Pi_{\pm ij} -2\kappa \Pi_{\pm ia} \hat\theta^{ab}_{\mp}\Pi_{\pm bj}
\end{array}\right )\, ,
\end{equation}
where ${\hat \theta}_{ \pm}^{ab}$ has been defined in (\ref{eq:bfc}). Details regarding derivation of the equations (\ref{eq:pipm}) and
(\ref{eq:pipmr}) are given in Ref.\cite{sazdam}. Reading the block components we obtained the NS-NS T-dual background fields in the flat background after dualization along  directions $x^a, \, (a=0,1, \cdots , d-1)$
\begin{eqnarray}\label{eq:tdualf}
& {}_a \Pi_{ \pm}^{ab} =   \frac{\kappa}{2} {\hat \theta}_{ \mp}^{ab}  \, ,  \qquad \qquad        & {}_a \Pi_{ \pm}^a{}_i =   \kappa {\hat \theta}_{ \mp}^{ab} \Pi_{ \pm bi}   \, ,   \\
& {}_a  \Pi_{ \pm i}{}^a =  -\kappa \Pi_{ \pm ib} {\hat \theta}_{ \mp}^{ba}  \, ,       & {}_a \Pi_{ \pm ij} = \Pi_{ \pm ij} -2\kappa \Pi_{ \pm ia} \hat\theta^{ab}_{ \mp}\Pi_{ \pm bj} \,  .
\end{eqnarray}
These are just the equations (\ref{eq:3})-(\ref{eq:1}).
The symmetric and antisymmetric parts of these expressions are T-dual metric and T-dual Kalb-Ramond field, which are in full agreement with the Refs.\cite{nasnpb,DNS2}.

\subsection{T-dual NS-R background fields $\Psi^\alpha_\mu$, $\bar\Psi^\alpha_\mu$}

Let us find the form of T-dual NS-R background fields, ${}_a \Psi^{\alpha a}$, ${}_a \Psi^\alpha_i$, ${}_a \bar\Psi^{\alpha a}$ and ${}_a \bar\Psi^\alpha_i$.
The T-dual current ${}_a J_{\pm M}$ (\ref{eq:parTJ}) should have the same form as initial one (\ref{eq:currentJ}) but in terms of the T-dual background fields
\begin{equation}\label{eq:aJJ}
\left(
\begin{array}{c}
2({}_a \Pi_\pm \;{}_a G^{-1})^a{}_b \;({}_a J)^b_{\pm}+2({}_a \Pi_\pm\; {}_a G^{-1})^{ai}\; ({}_a J)_{\pm i}\\
2({}_a \Pi_\pm\; {}_a G^{-1})_{ia}\; ({}_a J)^a_{\pm}+2({}_a \Pi_\pm\; {}_a G^{-1})_i{}^j\; ({}_a J)_{\pm j}\\
-({}_a G^{-1})_{ab}\; ({}_a J)_{\pm}^b-({}_a G^{-1})_a{}^i \;({}_a J)_{\pm i}\\
-({}_a G^{-1})^i{}_a\; ({}_a J)_{\pm}^a-({}_a G^{-1})^{ij}\; ({}_a J)_{\pm j}
\end{array}\right)=\left(
\begin{array}{c}
-(G^{-1})^{a\mu}J_{\pm \mu}\\
2(\Pi_\pm G^{-1})_i{}^\mu J_{\pm \mu}\\
2(\Pi_\pm G^{-1})_a{}^\mu J_{\pm \mu}\\
-(G^{-1})^{i\mu} J_{\pm \mu}
\end{array}\right).
\end{equation}
On the left-hand side of this equation we split the index $\mu$ in $a$ and $i$ components because in T-dual picture index $a$ has different position, it is now up. T-dual currents are written between the brackets
to make distinction between left subscript $a$ marking partial T-dualization and summation indices in the subspace spanned by $x^a$.

The information about T-dual NS-R background fields we can obtain from the lower $D$ components of the above equation.
In order to find the solution of these equations it is more practical to rewrite them using block-wise form of matrices given in Appendix and Ref.\cite{sazdam}
\begin{eqnarray}
&{}&-\tilde g_{ab} \;({}_a J)^b_\pm+2(\beta_1)_a{}^i\; ({}_a J)_{\pm i}=2(\tilde \beta\pm \frac{1}{2})_a{}^b J_{\pm b}+2(\beta_1)_a{}^i J_{\pm i}\, ,\nonumber\\
&{}&-2(\beta_1^T)^i{}_b\; ({}_a J)^b_\pm+\bar\gamma^{ij}\; ({}_a J)_{\pm j}=\gamma^{ia}\; J_{\pm a}+\bar\gamma^{ij} J_{\pm j}\, .
\end{eqnarray}
From the Eq.(3.18) of \cite{sazdam}
\begin{equation}
({}_a G^{-1})_{\mu\nu}=\left(
\begin{array}{cc}
g_{ab} & -2(BG^{-1})_a{}^j\\
2(G^{-1}B)^i{}_b & (G^{-1})^{ij}
\end{array}\right)=\left(
\begin{array}{cc}
 \tilde g & -2\beta_1\\
 -2\beta_1^T & \bar\gamma
\end{array}\right)\, ,
\end{equation}
(\ref{eq:gnmj}) and (\ref{eq:bgnmj}), we find the components of ${}_a G^{-1}$, $G^{-1}$ and $B G^{-1}$, respectively. In the first equation on right-hand side for $(\Pi_\pm G^{-1})_a{}^i$ stands just $(\beta_1)_a{}^i$ because $\delta_a{}^i=0$.

The difference
\begin{equation}\label{eq:veza2}
({}_a J)_{\pm i}-J_{\pm i}=(\bar\gamma^{-1})_{ij}\left[\gamma^{ja} J_{\pm a}+2(\beta_1^T)^j{}_b \;({}_a J)^b_{\pm}\right]\, ,
\end{equation}
obtained from the second equation, we put in the first equation which produces
\begin{equation}
2\left[\left(\tilde\beta\pm \frac{1}{2}\right)-\beta_1 \bar\gamma^{-1}\gamma\right]_a{}^b J_{\pm b}=-\left(\tilde g-4\beta_1 \bar\gamma^{-1}\beta_1^T\right)_{ab}\;{}_a J^b_\pm\, .
\end{equation}
From the definition of quantity $A_{ab}$ (\ref{eq:velA})
we get
\begin{equation}\label{eq:veza1}
({}_a J)^b_{\pm}=2\left[-A^{-1}(\tilde\beta\pm\frac{1}{2})+A^{-1}\beta_1 \bar\gamma^{-1} \gamma\right]^{bc} J_{\pm c}\, .
\end{equation}
Using the expression $A_{ab}=\hat g_{ab}$ (proved in \cite{sazdam}) and the relation (\ref{eq:usr}), we recognize $ab$ block-component of the relation (\ref{eq:pipm}). So, with the help of (\ref{eq:pipmr}) it is easily to see that
\begin{equation}
({}_a J)^b_\pm=2\;{}_a \Pi^{bc}_{\mp}J_{\pm c}=\kappa \;\hat\theta^{bc}_\pm J_{\pm c}\, .
\end{equation}

Note that now  the T-dual current ${}_a J_\pm^{\hat \mu }$ is of the form
\begin{equation}\label{eq:strujaJmut1}
{}_a J_\pm^{\hat \mu}=\pm\frac{2}{\kappa} \, {}_a \Psi^{\alpha \hat \mu }_{\pm } \, \, {}_a \pi_{\pm \alpha}\, ,
\end{equation}
where
\begin{equation}
{}_a \Psi^{\alpha \hat \mu }_{+} \equiv {}_a \Psi^{\alpha \hat \mu}\, ,\quad {}_a \Psi^{\alpha \hat \mu}_{-}\equiv {}_a \bar\Psi^{\alpha \hat \mu} \, ,\quad {}_a \pi_{+\alpha}\equiv  \pi_\alpha\, ,\quad {}_a \pi_{-\alpha}\equiv {}^\bullet \bar \pi_\alpha\, ,
\end{equation}
and as before
\begin{equation}\label{eq:strujea}
 J_{\pm \mu}=\pm\frac{2}{\kappa} \Psi^{\alpha}_{\pm \mu} \pi_{\pm \alpha}\, .
\end{equation}
So, the $a$ components of the T-dual NS-R fields are of the form
\begin{equation}\label{eq:apsi}
{}_a \Psi^{\alpha a}=\kappa \hat\theta^{ab}_+ \Psi_{b}^\alpha\, ,\quad {}_a \bar\Psi^{\alpha a}=\kappa \, {}_a \Omega^\alpha {}_\beta  \hat\theta^{ab}_- \bar\Psi_{b}^\beta \, .
\end{equation}

Substituting (\ref{eq:veza1}) into (\ref{eq:veza2}) we obtain
\begin{equation}\label{eq:vezai}
({}_a J)_{\pm i}-J_{\pm i}=\left(\bar\gamma^{-1}+4\bar\gamma^{-1}\beta_1^TA^{-1}\beta_1\bar\gamma^{-1}\right)_{ij}\gamma^{jb}J_{\pm b}-4\left[\bar\gamma^{-1}\beta_1^T A^{-1}(\tilde\beta\pm\frac{1}{2})\right]_i{}^a J_{\pm a}\,.
\end{equation}
With the help of (\ref{eq:DD})
the relation (\ref{eq:vezai}) transforms into
\begin{equation}
({}_a J)_{\pm i}-J_{\pm i}=2\left[\frac{1}{2}D^{-1}\gamma-2\bar\gamma^{-1}\beta_1^TA^{-1}(\tilde \beta\pm\frac{1}{2})\right]_i{}^a J_{\pm a}\, .
\end{equation}
From ${}_i{}^a$ component of (\ref{eq:pipm}) and (\ref{eq:pipmr}) we finally have
\begin{equation}
({}_a J)_{\pm i}=J_{\pm i}-2\kappa \Pi_{\mp i b} \hat\theta^{ba}_\pm J_{\pm a}\, .
\end{equation}
As in the previous case, using the expressions for currents (\ref{eq:strujaJmut1}) and (\ref{eq:strujea}), the final form of T-dual fields is
\begin{equation}\label{eq:ipsi}
{}_a \Psi^\alpha_i=\Psi^\alpha_i-2\kappa \Pi_{- ib}\hat\theta^{ba}_+ \Psi^\alpha_a\, ,\quad {}_a \bar\Psi^\alpha_i = {}_a \Omega^\alpha {}_\beta  (  \bar\Psi^\beta_i-2\kappa \Pi_{+ ib}\hat\theta^{ba}_- \bar\Psi^\beta_a )\, .
\end{equation}
The relations (\ref{eq:apsi}) and (\ref{eq:ipsi}) are in full agreement with the results from Ref.\cite{nasnpb} given by Eqs.(\ref{eq:apartpsi}) and (\ref{eq:partpsi}).

The upper $D$ components of Eq.(\ref{eq:aJJ}) produce the same result for T-dual background fields.
%%%%%%%%%%%%%%%%%%%%%%%%%%%%%%%%%%%%%%%%%%%%%%%%%%%%%%%%%%%%%%%%%%%%%%%%%%%%%%

\subsection{T-dual R-R field strength $F^{\alpha\beta}$}

Using the relations ${}_a{\cal H}={\cal T}^a {\cal H}{\cal T}^a$ and ${}_a J_{\pm}={\cal T}^a J_{\pm}$ we obtained the form of NS-NS and NS-R T-dual background fields of type II superstring theory. But we know from Buscher T-dualization procedure that T-dual R-R field strength ${}_a F^{\alpha\beta}$ has the form given in Eq.(\ref{eq:dualFalfabeta}). In this subsection we will derive this relation within the double space framework.

The R-R field strength $F^{\alpha\beta}$ appears in the action (\ref{eq:SB}) coupled
with fermionic momenta $\pi_\alpha$ and $\bar \pi_\alpha$ along which we do not
perform T-dualization. So, we did not double these variables. It is an analogue of $ij$-term in approach of Refs.\cite{Hull,Hull2} where $x^i$ coordinates are not doubled.
Consequently, as in \cite{Hull,Hull2} we should make some assumptions. Let us suppose that fermionic term $L (\pi_\alpha,\bar \pi_\alpha)$
is symmetric under exchange of R-R field strength $F^{\alpha\beta}$ with its T-dual ${}_a F^{\alpha\beta}$
\begin{equation}
 L = e^{\frac{\Phi}{2}} \, \pi_\alpha \, F^{\alpha\beta}\bar \pi_\beta  + e^{\frac{{}_a\Phi}{2}} {}_a \pi_\alpha \,\, {}_a F^{\alpha\beta}  \,\, {}_a \bar \pi_\beta \equiv   {\mathcal L} + {}_a  {\mathcal L}  \,  ,
\end{equation}
for some $F^{\alpha\beta}$ and ${}_a F^{\alpha\beta}$. This term should be invariant under T-dual transformation
\begin{equation}\label{eq:deltaF}
{}_a  {\mathcal L}  = {\mathcal L}  + \Delta {\mathcal L} \, .
\end{equation}
Taking into account the fact that two successive T-dualization are identity transformation, we obtain from (\ref{eq:deltaF})
\begin{equation}
{\mathcal L}  = {}_a  {\mathcal L} + {}_a \Delta  {\mathcal L}  \, .
\end{equation}
Combining last two relations we get
\begin{equation}\label{eq:uslovdl}
{}_a \Delta  {\mathcal L}  = -  \Delta  {\mathcal L}  \, .
\end{equation}
If $\Delta  {\mathcal L}$ has a form   $\Delta  {\mathcal L}= \, \pi_\alpha \, \Delta^{\alpha\beta}\bar \pi_\beta$ and consequently  ${}_a \Delta  {\mathcal L}= {}_a \pi_\alpha \, {}_a \Delta^{\alpha\beta} {}_a \bar \pi_\beta$, then
with the help of the first relation  (\ref{eq:bulet1}) we obtain the condition for $\Delta^{\alpha\beta}$
\begin{equation}\label{eq:uslovd}
{}_a \Delta^{\alpha\beta} = - \, \Delta^{\alpha \gamma}  {}_a \Omega_\gamma {}^ \beta \, .
\end{equation}
So, we should find the combination of background fields with two upper spinor indices which under T-dualization transforms as in (\ref{eq:uslovd}).
Using the expression for NS-R fields (\ref{eq:apartpsi}) and the equation $({}_a\hat\theta_\pm)_{ab}=\frac{2}{\kappa}\Pi_{\mp ab}=\frac{1}{\kappa^2}(\hat\theta^{-1}_\pm)_{ab}$ [see T-dual of (\ref{eq:tdualf}) and (\ref{eq:inv})],
it is easy to check that there are $D$ different solutions
\begin{equation}
\Delta_d^{\alpha\beta}=c \Psi^\alpha_a \hat\theta_-^{ab} \bar\Psi^\beta_b\, ,
\end{equation}
where $d=1,2,\dots D$ and $c$ is arbitrary constant.
Consequently, when we T-dualize $d$ dimensions $x^a\;(a=0,1,\dots d-1)$, from   (\ref{eq:deltaF})  we can conclude that  the T-dual R-R field strength has the form
\begin{equation}
e^{\frac{{}_a\Phi}{2}} \, {}_a F^{\alpha\beta}= (e^{\frac{\Phi}{2}} F^{\alpha\gamma}+c \Psi^\alpha_a \hat\theta^{ab}_- \bar\Psi^\gamma_b ) {}_a \Omega_\gamma {}^\beta \, .
\end{equation}
For $c=4\kappa$ we obtain the agreement with the expression (\ref{eq:dualFalfabeta}). Note that the fermionic term $L_d(\pi_\alpha,\bar \pi_\alpha)$ depends on $d$, number of directions
along which we perform T-duality as well as in Ref.\cite{Hull,Hull2}.

%%%%%%%%%%%%%%%%%%%%%%%%%%%%%%%%%%%%%%%%%%%%%%%%%%%%%%%%%%%%%%%%%%%%%%%%%%%%%%

\section{Conclusion}
\setcounter{equation}{0}

In this article we showed that the new interpretation of bosonic T-dualization procedure in double space
formalism offered in \cite{sazdam,sazda} is also valid in the case of type II superstring theory. We used the ghost free action of type II
superstring theory in pure spinor formulation in the approximation of quadratic terms and constant
background fields. One can obtain this action from action (\ref{eq:VSG}), which could be considered as an expansion in powers of fermionic coordinates. In the first part of analysis we neglect all terms in the action
containing powers of $\theta^\alpha$ and $\bar\theta^\alpha$. This approximation
is justified by the fact that action is a result of an interative procedure in which every step comes out from the previous one. Later, when we discussed proper fermionic variables, taking higher power terms 
we restore supersymmetric
invariants ($\Pi_\pm^\mu$, $d_\alpha$, $\bar d_\alpha$) as variables instead of $\partial_\pm x^\mu$, $\pi_\alpha$ and $\bar\pi_\alpha$.

We introduced the double space coordinate $Z^M=(x^\mu,y_\mu)$ adding to all bosonic initial coordinates, $x^\mu$,
the T-dual ones, $y_\mu$. Then we rewrote the T-dual transformation laws (\ref{eq:xtdual}) in terms of double space
variables (\ref{eq:tdual}) introducing the generalized metric ${\cal H}_{MN}$ and the current $J_{\pm M}$.
The generalized metric depends only on the NS-NS background fields of the initial theory. The current $J_{\pm M}$ contains
fermionic momenta $\pi_\alpha$ and $\bar \pi_\alpha$ along which we do not make T-dualization and depends also on NS-R background fields.
The R-R background fields do not appear in T-dual transformation laws.

The coordinate index $\mu$ is split in $a=(0,1,\dots d-1)$ and $i=(d,d+1,\dots D-1)$, where index $a$
marks subsets of the initial and T-dual coordinates, $x^a$ and $y_a$, along which we make T-dualization.
T-dualization is realized as permutation of the subsets $x^a$ and $y_a$ in the double space coordinate $Z^M$.
The main demand is that T-dual double space coordinates ${}_a Z^M=({\cal T}^a)^M{}_N Z^N$
satisfy the transformation law of the same form as the initial coordinates $Z^M$. From this condition we found the
T-dual generalized metric ${}_a {\cal H}_{MN}$ and the T-dual current ${}_a J_{\pm M}$. Because the initial and T-dual
theory are physically equivalent, ${}_a {\cal H}_{MN}$ and ${}_a J_{\pm M}$ should have the same form as
initial ones, ${\cal H}$ and $J_{\pm M}$, but in terms of the T-dual background fields. It produces the form
of NS-NS and NS-R T-dual background fields in terms of the initial ones which are in full accordance with the
results obtained by Buscher T-dualization procedure \cite{nasnpb,englezi}.

The supersymmetry case is not a simple generalization of the bosonic one, but requires some new interesting steps. The origin of the problem
is different T-duality transformation of world-sheet chirality sectors. It produces two possible sets of vielbeins in the T-dual theory
with the same T-dual metric. These vielbeins are related by particular local Lorentz transformation which depends on T-duality
transformation and which determinant is $(-1)^d$, where $d$ is the number of T-dualized coordinates. So, when we T-dualization along
odd number of coordinates then such transformation contains parity transformation. Consistency of T-duality with supersymmetry demands changing one of
two spinor sectors. We redefine the bar spinor coordinates, ${}_a\bar\theta\to {}_a^\bullet \bar\theta^\alpha={}_a \Omega^\alpha{}_\beta \bar\theta^\beta$, and
variable ${}_a \bar \pi_\alpha$, ${}_a \bar \pi_\alpha\to {}_a^\bullet \bar \pi_\alpha={}_a\Omega_\alpha{}^\beta \bar \pi_\beta$.
As a consequence bar NS-R and R-R background field include ${}_a\Omega$ in their T-duality transformations. For odd number of coordinates $d$
along which T-dualization is performed, ${}_a\Omega$ changes the chirality of bar gravitino $\bar\Psi^\alpha_\mu$ and chirality condition for $F^{\alpha\beta}$.
We need it to relate type IIA and type IIB theories.

Transformation law (\ref{eq:tdual}) induces the consistency condition which can be
considered as equation of motion of the double space action (\ref{eq:exact}). It contains an arbitrary term depending on
undualized variables ${L}(\pi_\alpha,\bar \pi_\alpha)$.
This is analogy with the term $\partial_+ x^i \Pi_{+ij}\partial_- x^j$ in approach presented in Ref.\cite{Hull,Hull2}. So,
to obtain T-dual transformation of R-R field strength $F^{\alpha\beta}$ we should make some additional assumptions.
Supposing that term ${L}(\pi_\alpha,\bar \pi_\alpha)$ is T-dual invariant  and taking into account
that two successive T-dualizations act as identity operator, we found the form of T-dual R-R field strength up to one
arbitrary constant $c$. For $c=4\kappa$ we get the T-dual R-R field strength ${}_a F^{\alpha\beta}$ as in Buscher
procedure \cite{nasnpb}.

T-duality transformation of the R-R field strength $F^{\alpha\beta}$ has two
contributions in the form of square roots. The contribution of dilaton produces the term $\sqrt{|\prod_{i=1}^d G_{a_ia_i}|}$.
On the other hand contribution of spinorial representation of Lorentz transformation ${}_a\Omega$ contains the same expression without absolute value $i^d \sqrt{\prod_{i=1}^d G_{a_ia_i}}$.
Therefore, T-dual R-R field strength ${}_a F^{\alpha\beta}$, besides rational expression, contains the expression  $i^d \sqrt{sign(\prod_{i=1}^d G_{a_i a_i})}$ (\ref{eq:tdf}).
If we T-dualize along time-like direction ($G_{00} > 0$), the square root does not produces imaginary unit $i$ and not canceled the one in front of the square root. So, T-dualization  along time-like direction  
maps type II superstring theories to type $II^\star$ ones \cite{timelike}.

The successive T-dualizations make a group called T-duality group. In the case of type II superstring T-duality transformations are performed by the same matrices ${\cal T}^a$ as in the bosonic string case
\cite{sazdam,sazda}. Consequently, the corresponding T-duality group is the same.

If we want to find
T-dual transformation of $F^{\alpha\beta}$ without any assumptions, we should follow approach of \cite{sazdam,sazda} and, besides all bosonic coordinates
$x^\mu$, double also all fermionic variables $\pi_\alpha$ and $\bar \pi_\alpha$. In other words, besides bosonic T-duality we should also consider fermionic T-duality \cite{fTdual}.
%%%%%%%%%%%%%%%%%%%%%%%%%%%%%%%%%%%%%%%%%%%%%%%e
\appendix
%%%%%%%%%%%%%%%%%%%%%%%%%%%%%%%%%%%%%%%%%%%%%%%
\cleq

\section{Block-wise expressions for background fields}\label{sec:dodatak}

In order to simplify notation we will introduce notations for
component fields following Ref.\cite{sazdam}.

For block-wise matrices there is a rule for inversion
\begin{equation}\label{eq:im}
\left(\begin{array}{cc}
A & B\\
C & D
\end{array}\right)^{-1}\,=
\left(
\begin{array}{cc}
(A-BD^{-1}C)^{-1}          &     -A^{-1}B(D-CA^{-1}B)^{-1}\\
-D^{-1}C(A-BD^{-1}C)^{-1}  &      (D-CA^{-1}B)^{-1}
\end{array}\right).
\nonumber\\
\end{equation}

For the metric tensor and the Kalb-Ramond background fields we define
\begin{equation}
G_{\mu \nu} = \left (
\begin{array}{cc}
{\tilde G}_{ab}    &    G_{aj}       \\
 G_{ib}            &   {\bar G}_{ij}
\end{array}\right )   \equiv
 \left (
\begin{array}{cc}
{\tilde G}    &    G^T       \\
 G            &   {\bar G}
\end{array}\right )   \, ,
\end{equation}
and
\begin{equation}
B_{\mu \nu} = \left (
\begin{array}{cc}
{\tilde b}_{ab}    &    b_{aj}       \\
 b_{ib}            &   {\bar b}_{ij}
\end{array}\right )   \equiv
 \left (
\begin{array}{cc}
{\tilde b}    &    -b^T       \\
 b            &   {\bar b}
\end{array}\right )   \, .
\end{equation}
We also define notation for inverse of the matric
\begin{equation}\label{eq:gnmj}
(G^{-1})^{\mu \nu} = \left (
\begin{array}{cc}
{\tilde \gamma}^{ab}    &    \gamma^{aj}       \\
 \gamma^{ib}            &   {\bar \gamma}^{ij}
\end{array}\right )   \equiv
 \left (
\begin{array}{cc}
{\tilde \gamma}    &    \gamma^T       \\
 \gamma          &   {\bar \gamma}
\end{array}\right )   \, ,
\end{equation}
and for the effective metric
\begin{equation}\label{eq:gdef}
G^E_{\mu \nu} = G_{\mu \nu} -4 B_{\mu\rho} (G^{-1})^{\rho \sigma} B_{\sigma \nu}
= \left (
\begin{array}{cc}
{\tilde g}_{ab}    &    g_{aj}       \\
 g_{ib}            &   {\bar g}_{ij}
\end{array}\right )   \equiv
 \left (
\begin{array}{cc}
{\tilde g}    &    g^T       \\
 g            &   {\bar g}
\end{array}\right )   \, .
\end{equation}

Note that because $G^{\mu \nu}$ is inverse of $G_{\mu \nu}$  we have
\begin{eqnarray}\label{eq:cgama}
& \gamma = - {\bar G}^{-1} G {\tilde \gamma} = - {\bar \gamma} G {\tilde G}^{-1} \, , \qquad
& \gamma^T = - {\tilde G}^{-1} G^T {\bar \gamma} = - {\tilde \gamma} G^T {\bar G}^{-1} \, , \nonumber \\
& {\tilde \gamma} = ({\tilde G}- G^T {\bar G}^{-1} G)^{-1} \,  ,
& {\bar \gamma} = ({\bar G}- G {\tilde G}^{-1} G^T)^{-1} \,  , \nonumber \\
&{\tilde G}^{-1}= {\tilde \gamma}- \gamma^T {\bar \gamma}^{-1} \gamma \, ,
&{\bar G}^{-1}= {\bar \gamma}- \gamma {\tilde \gamma}^{-1} \gamma^T \,  .
\end{eqnarray}

It is also useful to introduce new notation for expression
\begin{equation}\label{eq:bgnmj}
(BG^{-1})_\mu{}^\nu  = \left (
\begin{array}{cc}
{\tilde b}  {\tilde \gamma}- b^T \gamma      &    {\tilde b}\gamma^T- b^T {\bar \gamma}         \\
  b  {\tilde \gamma} + {\bar b} \gamma       &    b \gamma^T + {\bar b} {\bar \gamma}
\end{array}\right )   \equiv
 \left (
\begin{array}{cc}
{\tilde \beta}    &    \beta_1       \\
 \beta_2            &   {\bar \beta}
\end{array}\right )   \, .
\end{equation}

We denote by hat $\,\,{\hat {} }\,\,$  expressions similar to  the effective metric (\ref{eq:gdef}) and non-commutativity  parameters
 but with all contributions from $ab$ subspace
\begin{equation}\label{eq:ghat}
{\hat g}_{ab} = ({\tilde G} -4 {\tilde b}  {\tilde G}^{-1} {\tilde b})_{ab}   \, , \qquad
 {\hat \theta}^{ab} = -\frac{2}{\kappa} ({\hat g}^{-1} {\tilde b} {\tilde G}^{-1})^{ab}   \, .
\end{equation}
Note that ${\hat g}_{ab} \neq {\tilde g}_{ab}$ because ${\tilde g}_{ab}$ is projection of $g_{\mu \nu}$ on subspace $ab$.
It is extremely useful to introduce background field combinations
\begin{equation}\label{eq:bfc}
\Pi_{\pm ab}= B_{ab} \pm  \frac{1}{2} G_{ab}   \,  \qquad
{\hat \theta}^{ab}_\pm =  -\frac{2}{\kappa} ({\hat g}^{-1} {\tilde \Pi}_\pm {\tilde G}^{-1})^{ab} =
{\hat \theta}^{ab} \mp \frac{1}{\kappa}  ({\hat g}^{-1})^{ab} \, ,
\end{equation}
which are inverse to each other
\begin{equation}\label{eq:inv}
{\hat \theta}^{ac}_\pm  \Pi_{\mp cb} = \frac{1}{2 \kappa} \delta^a_b   \, .
\end{equation}

The quantity $A_{ab}$ is defined as
\begin{equation}\label{eq:velA}
A_{ab}=(\tilde g-4\beta_1 \bar\gamma^{-1}\beta_1^T)_{ab}\, .
\end{equation}

One can prove the relation \cite{sazdam}
\begin{eqnarray}\label{eq:usr}
({\tilde g}^{-1} \beta_1  D^{-1})^a {}_i= ({\hat g}^{-1} \beta_1 {\bar \gamma}^{-1})^a {}_i  \, ,
\end{eqnarray}
where $D^{ij}$ is defined in Eq.(3.21) of \cite{sazdam}
\begin{equation}\label{eq:DD}
D^{ij}=(\bar\gamma-4\beta_1^T \tilde g^{-1}\beta_1)^{ij}\, ,\quad (D^{-1})_{ij}=\left(\bar\gamma^{-1}+4\bar\gamma^{-1}\beta_1^TA^{-1}\beta_1\bar\gamma^{-1}\right)_{ij}\, .
\end{equation}
%%%%%%%%%%%%%%%%%%%%%%%%%%%%%%%%%%%%%%%%%%%%%%%%%

\end{document}